\documentclass[a4paper,11pt]{article}
\usepackage{graphicx}
\usepackage{caption}
\usepackage{subcaption}
\usepackage{jheppub}

\usepackage[utf8]{inputenc}
\usepackage[T1]{fontenc}
\usepackage{amsmath}
\usepackage{amsfonts}
\usepackage{amssymb}
\usepackage{graphicx}
\usepackage{braket}

\newcommand{\dd}{\mathrm{d}}
\newcommand{\avsum}{\sideset{}{^\prime}\sum}
\newcommand{\ms}{\mskip 1.5mu}
\newcommand{\tvec}[1]{\boldsymbol{#1}}

\newcommand{\mvec}[1]{\vec{\mskip 0.5mu #1}\mskip 1.5mu}
\newcommand{\tr}{\operatorname{tr}}

\newcommand{\half}{\tfrac{1}{2}}

\title{Double Parton Distributions in the nucleon:\\
   Lattice parameter dependence}

\author[a]{Daniel Reitinger,}
\author[b,c]{Christian Zimmermann}
\author[a]{and Andreas Sch{\"a}fer}

\affiliation[a]{Institute for Theoretical Physics, University of Regensburg,
   93040 Regensburg, Germany}
\affiliation[b]{Department of Physics and Astronomy, University of Kentucky,
   Lexington, KY 40506, USA}
\affiliation[c]{Nuclear Science Division, Lawrence Berkeley National Laboratory,
   Berkeley, CA 94720, USA}   
\begin{document}

\abstract{
To make full use of %of high luminosity and high energy colliders, especially 
the HL-LHC, especially in the search for BSM physics, all Standard Model backgrounds have to be understood with increased precision. This is especially difficult for Double Parton Interactions (DPIs) and more generally Multi Parton Interactions (MPIs), which are very difficult to isolate precisely in experiment. DPIs are parameterized by Double Parton Distributions (DPDs) and the lattice determination of the latter is, therefore, especially attractive. 
In our previous work on DPDs on the lattice, we focused on the results obtained from one gauge ensemble with higher than physical pion mass. In this work, we extend our efforts by investigating artifacts introduced by finite volume and the dependence on the lattice spacing. Moreover, we consider the dependence on the mass parameters. This study is carried out by employing the Wilson-Clover fermion gauge ensembles generated by the CLS collaboration.}

\emailAdd{Daniel.Reitinger@physik.uni-regensburg.de}
\maketitle

\tableofcontents

\section{Introduction}
If all goes according to plan high-energy particle physics will get a tremendous boost in the 2030's when more or less simultaneously the HL-LHC \cite{Apollinari:2015wtw} EIC \cite{Accardi:2012qut,AbdulKhalek:2021gbh} and EicC \cite{Anderle:2021wcy} will start operating. Of the very many research fields which will profit from this development we focus on MPIs i.e. interactions in which several pairs of partons in the same collision experience hard interactions. The QCD description of MPIs is highly non-trivial, but in view of their relevance for hadron structure and dynamics, in particular the identification of BSM signals, it is also indispensable. It could very well happen that experimental future data either heralds the advent of New Physics or the fact that some QCD MPI background is still not fully understood. Let us stress that their better understanding is also a research goal in its  own right. In practice all major event generators include features for modeling MPIs, see e.g. for Pythia \cite{Sjostrand:2017cdm,Bierlich:2022pfr,Fedkevych:2025lgp}, for Herwig \cite{Bellm:2019icn,Bellm:2019zci}, or for Sherpa \cite{Sherpa:2019gpd}. Another, more general improvement is e.g.  \cite{Cabouat:2020ssr}.

In principle, one would like to know the full multi-particle wave-function of all well-established hadrons with high precision. In view of the enormous amount of information contained in such functions even for limited precision this is hardly possible. Therefore, the field develops by evolution from simple to ever more complicated functions, from parton distribution functions (PDFs) and formfactors, to Generalized Parton Distributions (GPDs) and Transverse Momentum Dependent PDFs (TMDPDFs), and multi field correlators needed, e.g., to describe MPIs. The number of independent functions increases substantially with their complexity. For example the nucleon has three twist-2 quark PDFs per flavor, eight twist-2 quark TMDs per flavor, up to 116 DPDs per flavor combination. This simply reflects the fact that GPDs, TMDPDFs, certain distribution amplitudes (DAs), certain formfactors and especially DPDs  contain far more information than PDFs. So, this is a massive advantage rather than a disadvantage compared to traditional PDF calculations. On the back-side, however, it implies that the comprehensive determination
of, e.g., all nucleon DPDs is extremely demanding.

Because DPI amplitudes contain four rather than two parton fields and parton distributions increase strongly with decreasing $x$ the relative importance of DPIs for high energy collisions increases strongly with increasing collision energy. Therefore, most investigations of DPIs study high-energy hadron-hadron collisions \cite{UA2:1991apc,CDF:1993sbj,CDF:1997lmq,CDF:1997yfa,ATLAS:2012yve,CMS:2012qbp,CMS:2019jcb,CMS:2021qsn,CMS:2022pio,ATLAS:2025bcb}. Recently, however, it was argued that the large intensity of the EIC and its different event structure could compensate its lack of energy \cite{Blok:2026kxx,Butterworth:2024hvb,Ceccopieri:2021luf}. In view of the problems just sketched this is relevant because in particular SIDIS events contain complementary information and have different systematic errors than purely hadronic events. Things become even more difficult if one turns to p+A collisions, especially ultra-peripheral collisions, where one of the colliding nuclei acts as source of, e.g., photons which interact with the other nucleus and for which a large research program is realized at the LHC, see e.g., \cite{Baltz:2007kq,ALICE:2026mlg,Krauss:2026dnq,ATLAS:2026cxr,Wang:2026fvy} and citations given there.
%shares e.g. shadowing and anti-shadowing effects with electron nucleus scattering and for which a characteristic scaling with nucleon number was observed. 
In \cite{LHCb:2020jse} a strong enhancement of DPIs compared to Single Parton Interactions (SPIs) was reported for such proton-lead collisions compared to pp collisions, see also \cite{Koshkarev:2022mgi,CMS:2024wgu}.

There exist quite a few non-perturbative effects, e.g. shadowing and antishadowing effects, which on the one hand make DPIs difficult to interpret in p+A, but on the other hand adds even more physics information, e.g. on many-body effects in QCD. A very recent example is the relevance of resonance decay recently claimed by ALICE \cite{ALICE:2025byl}. All of these examples illustrate the complexity of this research field.
This complexity probably explains why some groups focus on especially simple reaction channels like "same-sign W pair production" \cite{Kom:2011bd,Kom:2011nu,Borschensky:2016nkv,Fedkevych:2020cmd}. They also convey a feeling for the great achievement of, e.g., the ATLAS and CMS experiment which established non-zero DPI contributions for LHC p+p events \cite{ATLAS:2025bcb,CMS:2022pio} in spite of all the sketched complications. 

To reduce the complexity of the description of DPIs, most analyses are based on the "pocket formula" 
\begin{equation}
    \sigma_{DPS}=\frac{1}{1+\delta_{AB}}\frac{\sigma_A \sigma_b}{\sigma_{\rm eff}} 
\label{poket}
\end{equation}
which is known to be only approximately correct, see Figure 5 in \cite{ATLAS:2025bcb} and keep in mind that the factor $\sigma_{\rm eff}$ enters in the denominator such that it cannot be too small. Much effort has been invested in recent years to improve the treatment of MPIs both in Monte Carlo generators, see e.g. \cite{Fedkevych:2022myf} and references given there, and by implementing more QCD features such that our understanding of MPIs is bettered step by step.
Examples of such improvements are, e.g., QCD evolution of DPDs up to two loop splitting, see e.g. \cite{Diehl:2023jje,Diehl:2026imf}, constraints from energy and angular momentum sumrules \cite{Diehl:2020xyg}, treatment of color and flavor interference effects, which goes beyond the factorization assumption of Eq.(\ref{poket}) \cite{Blok:2022mtv,Diehl:2023jje}. 

Also many other aspects of MPI physics, as well as specific reaction channels were investigated, generating far too many publications for all to be cited here. So, we just mention some of the other most prominent ones \cite{Blok:2020ckm,Gaunt:2009re,Gaunt:2012tfk,diehl2017doublepartonscatteringtheory,Diehl:2017kgu,Diehl:2018kgr}. Despite all of these efforts and a tremendous amount of work it is probably fair to say that the experimental study of MPIs has not yet reached the status of precision physics. The first sentence in the abstract of \cite{Andersen:2023hzm} summarizes the situation quite concisely in that MPIs are fascinating, nearly omnipresent phenomena, which are "remarkably difficult to study quantitatively". 

In this situation it is natural to turn to lattice QCD as a source of complementary data, which
is what we did in \cite{Bali:2021gel,Bali:2020mij,Bali:2018nde,Reitinger:2024ulw,Diehl:2025kcr} and do in the present work, 
but many more lattice simulations would be needed, also for non-trivial kinematics, especially large hadron momenta $p^{\mu}$, for which signal/noise tends to be small, to get a comprehensive picture. With only traditional, Mellin momenta based methods this is unfeasible. 
Consequently, in the present paper we focus on zero momentum kinematics $\vec p=\vec 0$ and the dependence on lattice spacing $a$ and quark mass. Not surprisingly we find that different channels and flavor combinations, each resulting in a sum of several amplitudes, show quite different behavior. Some depend strongly on $a$ or quark mass or both and some are basically independent of them, especially those given by vector-vector current correlations. (The analysis of both dependencies is the main advance compared to our earlier work.) 

In addition to the computational problems just sketched, exotic correlators tend to be small such that their experimental isolation is often especially demanding. This is even more so because disentangling the mixing with higher-twist contributions also becomes ever more difficult with increasing complexity. The upshot of all of this is that for the precise lattice determination of DPDs fundamentally improved lattice QCD techniques are needed and all available information has to be combined, including input from lattice QCD and future experiments.

Just like the experimental investigation of DPIs is much more difficult than that of Single Parton Interactions (SPIs), this is also true for lattice QCD simulations. This starts from the facts that to describe DPIs one has to determine also 4-point functions on the lattice rather than only 2-point and 3-point functions and that DPDs depend on additional variables such that their determination requires much more computer resources. Still, over the years, we have made significant progress towards this goal. This work and \cite{Diehl:2025kcr} present the latest developments, where the focus of both papers is quite different. While the present paper focuses on the precision already reached in our lattice simulations for the needed invariant functions and hardly dependent on any model assumptions,  \cite{Diehl:2025kcr} investigates how strongly these results constrain specific DPD models. At present, these models are primarily constrained by our lattice data and some analytic facts which is insufficient to reach high precision.   
However, at present we have only rather little and somewhat controversial experimental information. The new accelerators, which will very significantly improve the investigation of internal hadron structure, and dynamics, should change this situation, allowing for joint fits of experimental and lattice data which significantly reduce the freedom of modeling.  

Besides the sketched planned improvement of the experimental situation, there are also several theoretical developments which we expect to change the situation: 
The standard formalism to extract information on, e.g., PDFs, DAs, TMDs, GPDs, etc. as well as DPDs is based on operator product expansion (OPE) and allows, e.g.,  to determine Mellin moments of PDFs as well as DPDs. Until recently, this was only possible for a few leading Mellin moments (like 2, 3, or 4) due to operator mixing following from the fact that the symmetry group of Euclidean lattice QCD is H(4), which is much smaller than that of continuum QCD, i.e. O(4). Consequently, irreducible representations of O(4) become reducible representations of H(4) leading to operator mixing which is especially problematic if it results in mixing with lower dimensional operators and thus factors proportional to inverse powers of the lattice spacing $a$, see \cite{Kronfeld:1984zv,Martinelli:1987zd,Beccarini:1995iv,Gockeler:1996mu}. 

Recently, however, it was argued in \cite{Shindler:2023xpd} (building on \cite{Davoudi:2012ya}) that using gradient flow methods (and thus a different regularization scheme) these problems can be circumvented such that the precise determinations of higher Mellin moments become feasible. If practical numerical applications confirm this expectation, future lattice studies of DPDs will allow to determine more parameters. In  \cite{Shindler:2023xpd}  it is also considered whether this method could also work for quasi-distributions without reaching a definitive conclusion. If it does this is quite interesting in connection with LaMET calculations and \cite{Zhang:2023wea,Jaarsma:2023woo} where methods to calculate the full functional form of DPDs directly from LaMET are proposed.
%While all of these developments provide strong motivation for future lattice QCD work on DPDs, the strongest one would definitely the observation signals for tantalizing BSM physics will be identified would depend on the exclusion of a potential MPI based QCD explanation the motivation for investing such resources might be given. 

Another recent development is kinematically enhanced correlators which allow to greatly improve signal/noise for hadrons with large $\vec p$, see \cite{Zhang:2025hyo}. As using kinematically improved correlators has very significant numerical advantages, it became very rapidly a standard element of lattice simulations see, e.g., \cite{Holligan:2025baj,Tan:2025ofx,CLQCD:2025dod,Ji:2026vir,Gao:2026hix,Gao:2026wlz,LPC:2026vyv,Zhang:2026lle,Grebe:2026qmt,LPC:2026lcj,LPC:2026mvw,Avkhadiev:2026xyf,Constantinou:2026wii} and references given there. Our recent contribution to this development is \cite{Reitinger:2026hta}.

Still, there is the problem of the sheer number of parameters and thus required lattice simulations. Also for this, there might be a solution in sight. This problem is already encountered for, e.g., LaMET calculations of TMDs, though in much milder form. A comprehensive analysis, as in \cite{LatticeParton:2024mxp}, requires hundreds of independent runs and fits to the obtained data. While all of these are somewhat repetitive their automatization was so far not possible. This has recently changed, for example with the development of the AI tool LQCDM{\sc aster}, see \cite{Gao:2026cpd} co-authored by Qi'an Zhang and Wei Wang who are senior members of the Lattice Parton Collaboration (LPC) to which one of us (AS) also belongs. 

It is quite improbable that all of these ideas will come to fruition, but there will be others. These are just meant as illustrations that progress beyond traditional lattice techniques is possible and even probable such that the lattice determination of DPDs could become a standard procedure in the mid-term future.

{The analysis of our previous works on DPDs was mainly built upon the results for one lattice, i.e.\ for one fixed volume and lattice spacing, as well as one higher than physical pion mass \cite{Bali:2021gel,Bali:2020mij,Bali:2018nde,Reitinger:2024ulw,Diehl:2025kcr}. The current work attempts to analyze the impact of those lattice parameters on observables related to DPDs. To this end, we extend our previous simulations by including further gauge ensembles. Again, we employ the Wilson-clover gauge ensembles generated by the CLS collaboration \cite{Bruno:2014jqa}.}

The paper is organized as follows: sections~\ref{sec:theory} and~\ref{sec:lattice-calc} provide a brief overview of the theoretical framework of double parton distributions (DPDs) and the lattice methodology. Section~\ref{sec:results} presents the results of our analysis for lattice artifacts in section~\ref{sec:artifacts}, lattice spacing dependence in section~\ref{sec:adep} and pion-mass dependence in section~\ref{sec:mdep}. Our findings are summarized in section~\ref{sec:summary}, which also provides an outlook on future work. %The simulation parameters are listed in the respective sections discussing each analysis.

\section{Theory background}
\label{sec:theory}

In this section, we review basic definitions and properties of DPDs, including
their extension to nonzero skewness $\zeta$ introduced in our previous
lattice studies \cite{Bali:2020mij,Bali:2021gel}. We express all four-vectors $v^\mu$ in light cone coordinates $v^{ \pm}=\left(v^0 \pm v^3\right) / \sqrt{2}$ and $ \boldsymbol{v}=\left(v^1, v^2\right)$.

\subsection{DPDs, skewed DPDs, and their Mellin moments}

DPDs are defined as matrix elements of the product of two twist-two color singlet operators\footnote{As discussed in \cite{Diehl_2012}, there are also color octet contributions, which we do not consider in this work.}. The unpolarized, i.e.\ spin-averaged, DPDs of two quarks in a proton are defined in a frame where the proton has no transverse momentum, i.e. $\tvec{p}=0$:

\begin{align}
   \label{eq:DPD}
   F_{a_1 a_2}\left(x_1, x_2, \zeta, \tvec{y}\right)
   =
   2 p^{+} \int \mathrm{d} y^{-}
   & 
   e^{-i \zeta y^{-} p^{+}}
   \int \frac{\mathrm{d} z_1^{-}}{2 \pi}
   \frac{\mathrm{~d} z_2^{-}}{2 \pi} \,
   e^{i \left(x_1^{} z_1^{-} + x_2^{} z_2^{-}\right) p^{+}}
   \nonumber \\[0.2em]
   &
   \times \sideset{}{^{\prime}}\sum_{\lambda}
   \bra{p,\lambda}
   \mathcal{O}_{a_1}(y, z_1) \,
   \mathcal{O}_{a_2}(0, z_2)
   \ket{p,\lambda}
   \,.
\end{align}
where $a_{1,2}$ indicate the quark flavor and polarization, where $q$, $\Delta q$ and $\delta q$ correspond to a quark of flavor $q$ which is respectively unpolarized, longitudinally polarized or transversely polarized. The averaging over proton polarization $\lambda$ is denoted as $\sum_{\lambda}^{\prime}=\frac{1}{2} \sum_\lambda$. 
The twist-two operators in \eqref{eq:DPD} read

\begin{align}
   \label{eq:qqbar-ops}
   \mathcal{O}_a(y, z)
   &=
   \left.
   \bar{q}\! \left(y - \frac{1}{2} z\right)
   \Gamma_a \,
   q\! \left(y + \frac{1}{2} z\right)
   \right|_{z^{+}=y^{+}=0,\, \tvec{z} = \tvec{0}}\,,
\end{align}
with

\begin{align}
    \Gamma_q = \frac{1}{2} \gamma^+ \,,
    \qquad
    \Gamma_{\Delta q} = \frac{1}{2} \gamma^+ \gamma_5 \,,
    \qquad
    \Gamma^j_{\delta q} = \frac{i}{2} \sigma^{j+} \gamma_5
\end{align}
Notice that, strictly speaking, this expression holds only in light cone gauge, otherwise a Wilson line has to be inserted between  $y - \half z$ and $y + \half z$.
%
%The operator $\mathcal{O}_{g}$ for gluons can for instance be found in Section~2.2 of \cite{Diehl:2011yj}. 
UV divergences are understood to be renormalized in the $\overline{\text{MS}}$ scheme. We suppress the notation of scale dependence in operators and matrix elements in the following.

If $\zeta = 0$, $x_i$ are the respective \textit{momentum fractions} of the two partons w.r.t.\ $p^+$. $\zeta$ is the so-called \textit{skewness} parameter, which introduces a difference in the momentum fraction of the emitted and absorbed quark in the proton state and its complex conjugate, so that the corresponding momentum fractions read $x_i-\zeta/2$ or $x_i+\zeta/2$, respectively. In the cases relevant for double parton scattering, $\zeta$ equals zero. As we discuss in section \ref{ssec:tcme}, in the context of our lattice calculation, we rather consider the Fourier transform w.r.t.\ $\zeta$, where the Fourier conjugate variable is the so-called Ioffe time $\omega = p^+ y^- = p\cdot y$.

\begin{figure}
\begin{center}
\subfloat[\label{support-dpd} $x_i \pm \zeta / 2 > 0$]{
\includegraphics[width=0.49\textwidth]{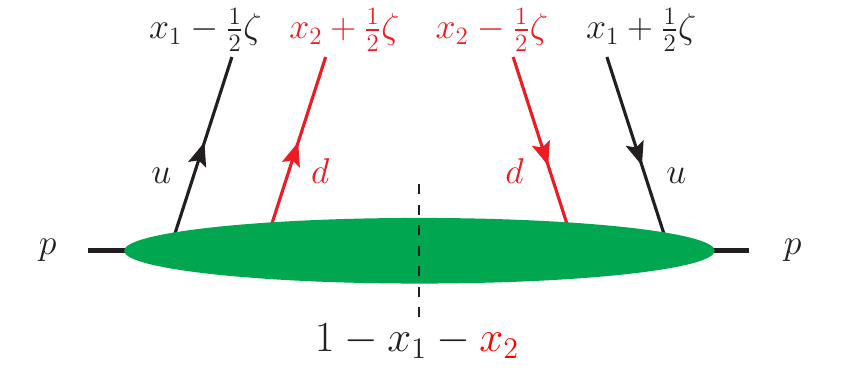}
}
\subfloat[\label{support-central} $\zeta / 2 \pm x_i > 0$]{
\includegraphics[width=0.49\textwidth]{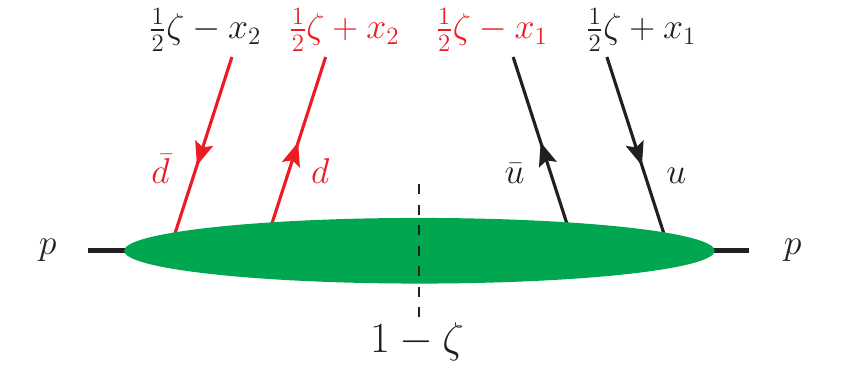}
}
\caption{Illustration of the skewed DPD $F_{u d}\left(x_1, x_2, \zeta,
y^2\right)$ for the cases where (a) all fractions $x_i \pm \zeta / 2$ or (b) all
fractions $\zeta / 2 \pm x_i$ are positive, taken from \cite{Diehl:2025kcr}.  Below the vertical dashed line we
give the total momentum fraction carried by the spectator partons.}
\label{zeta-distrib}
\end{center}
\end{figure}
%
%The skewness parameter $\zeta$ introduces a difference w.r.t.\ momentum fractions between emitted and absorbed partons, again potentially altering their interpretation as partons or anti-partons. 
The sign of the momentum fraction $x_i \pm \zeta/2$ impacts the interpretation of whether the considered partons correspond to quarks or anti-quarks. If all momentum fractions are positive, we have two quarks in the wave function and its complex conjugate. A change in sign of the momentum fraction $x_i \pm \zeta/2$ transforms the corresponding quark into an antiquark in the complex conjugate wave function and vice versa. This is illustrated in Figure \ref{support-central}, where an emitted $u$-quark became an absorbed $\bar{u}$-quark and an absorbed $d$-quark turned into an emitted $\bar{d}$-quark since $x_i<\zeta/2$. A detailed discussion of the different regions of $x_i$ and $\zeta$ and their interpretation is given in \cite{Bali:2020mij,Bali:2021gel}.

Rotational symmetry w.r.t.\ the transverse plane allows us to decompose the DPDs $F(x_1,x_2,\zeta,\tvec{y})$ in terms of rotational invariant functions $f(x_1,x_2,\zeta,y^2)$, where $y^2 = -\tvec{y}^2$, since $y^+ = 0$. At leading twist, we have six independent functions: $f_{qq^\prime}$, $f_{\Delta q \Delta q^\prime}$, $f_{\delta q q^\prime}$, $f_{q \delta q^\prime}$, $f_{\delta q \delta q^\prime}$, and $f^t_{\delta q \delta q^\prime}$. For details on the decomposition, see \cite{Bali:2021gel}.

The formalism can be extended to flavor changing operators, i.e.:

\begin{align}
   %\label{eq:qqbar-ops}
   \mathcal{O}_a(y, z)
   &=
   \left.
   \bar{q}\! \left(y - \frac{1}{2} z\right)
   \Gamma_a \,
   q^\prime\! \left(y + \frac{1}{2} z\right)
   \right|_{z^{+}=y^{+}=0,\, \tvec{z} = \tvec{0}}\,,
\end{align}
with $q\neq q^\prime$, in which case denote $a_{1,2} = (q q^\prime),  \Delta(q q^\prime),  \delta(q q^\prime)$, following the notation of \cite{Reitinger:2024ulw}. The corresponding functions $F_{a_1 a_2}$ correspond to flavor interference effects in double parton scattering and are therefore called flavor interference distributions or flavor interference DPDs.

In the context of lattice simulations, it is useful to introduce Mellin moments of DPDs w.r.t.\ the momentum fractions $x_i$:

\begin{align}
  \label{eq:skewed-inv-mellin-mom-def}
   I_{a_1 a_2}(\zeta, y^2)
   &=
   \int_{-1}^{1} \dd x_1^{} \int_{-1}^{1} \dd x_2^{} \;
   f_{a_1 a_2}(x_1,x_2,\zeta, y^2)
   \notag \\[0.2em]
   &=
   \frac{2}{(p^{+})^2} \int \mathrm{d} \omega \, e^{-i \zeta \omega} \,
   \sideset{}{^{\prime}}\sum_{\lambda}
   \bra{p,\lambda}
   \mathcal{O}_{a_1}\left(y, 0\right)
   \mathcal{O}_{a_2}\left(0, 0\right)
   \ket{p,\lambda}
   \,,
\end{align}
In analogy, define the Mellin moment $I^t_{\delta q \delta q^\prime}$ for the DPD $f^t_{\delta q \delta q^\prime}$.
The integration over the momentum fractions $x_i$ transforms the light-cone operators in \eqref{eq:DPD} into local operators, which enables us to evaluate the corresponding matrix element directly on a lattice with Euclidean time. This approach has already been used in \cite{Bali:2021gel,Reitinger:2024ulw}. An alternative approach has been proposed in the context of large momentum effective theory (LaMET) \cite{Jaarsma:2023woo,Zhang:2023wea}.

%%%%%%%%%%%%%%%%%%%%%%%%%%%%%%%%%%%%%%%%

\subsection{Two-current matrix elements}
\label{ssec:tcme}

We review some properties and relations regarding generic nucleon matrix elements of two local currents, which are directly related to the Mellin moments of DPDs through eq.\ \eqref{eq:skewed-inv-mellin-mom-def}. Those matrix elements can be readily calculated on the lattice with Euclidean spacetime. We only give a brief summary of what has already been worked out in \cite{Bali:2021gel}.

First of all, we define:
    
\begin{align}
  \label{eq:mat-els}
M^{\mu_1 \cdots \mu_2 \cdots}_{q_1 q_2 q_3 q_4, i_1 i_2}(p,y)
&:= \avsum_\lambda \bra{p,\lambda} J^{\mu_1 \cdots}_{q_1 q_2, i_1}(y)\,
              J^{\mu_2 \cdots}_{q_3 q_4, i_2}(0) \ket{p,\lambda} \,,
\end{align}
where we average over the spin of the external nucleon states in analogy to \eqref{eq:DPD}, or \eqref{eq:skewed-inv-mellin-mom-def}, respectively. The currents $J$ in \eqref{eq:mat-els} are local quark bilinears defined as:

\begin{align}
  \label{eq:local-ops}
J_{qq^\prime, V}^\mu(y) &= \bar{q}(y) \ms \gamma^\mu\ms q^\prime(y) \,,
&
J_{qq^\prime, A}^\mu(y) &= \bar{q}(y) \ms \gamma^\mu \gamma_5\, q^\prime(y) \,,
&
J_{qq^\prime, T}^{\mu\nu}(y) &= \bar{q}(y) \ms \sigma^{\mu\nu} \ms q^\prime(y) \,.
\end{align}    
For space-like separations $y$, the operators \eqref{eq:local-ops} commute. On a Euclidean lattice, this is trivially the case and leads to the following relation:

\begin{align}
\label{eq:dpd_sym}
    M^{q_1 q_2 q_3 q_4}_{ij}(p, y) = M^{q_3 q_4 q_1 q_2}_{ji}(p, -y) \,.
\end{align}
This allows us to average over data of the type $M^{q_1 q_2 q_1 q_2}_{VT}$, thus increasing statistics for these cases or to obtain data for interchanged flavors.
In order to make contact with the DPDs, we decompose the matrix elements in terms of Lorentz invariant functions:

\begin{align}
  \label{eq:tensor-decomp}
M^{\{\mu\nu\}}_{q_1 q_2 q_3 q_4, V V}
        - \tfrac{1}{4} g^{\mu\nu} g_{\alpha\beta}
                M^{\alpha\beta}_{q_1 q_2 q_3 q_4, V V}
 & = u_{V V,A}^{\mu\nu}\, A_{(q_1 q_2) (q_3 q_4)}^{}
   + u_{V V,B}^{\mu\nu}\, m^2\ms B_{(q_1 q_2) (q_3 q_4)}^{}
\nonumber \\[0.1em]
 &\quad + u_{V V,C}^{\mu\nu}\, m^4\ms C_{(q_1 q_2) (q_3 q_4)}^{} \,,
\nonumber \\[0.1em]
M^{\mu\nu\rho}_{q_1 q_2 q_3 q_4, T V} + \tfrac{2}{3} g_{\vphantom{q_1T}}^{\rho[\mu} M^{\nu]\alpha\beta}_{q_1 q_2 q_3 q_4, T V} g_{\alpha\beta}
 & = u_{T V,A}^{\mu\nu\rho}\, m\ms A_{\delta (q_1 q_2) (q_3 q_4)}^{}
   + u_{T V,B}^{\mu\nu\rho}\,  m^3\ms B_{\delta (q_1 q_2) (q_3 q_4)}^{} \,,
\nonumber \\[0.3em]
\tfrac{1}{2} \, \bigl[ M^{\mu\nu\rho\sigma}_{q_1 q_2 q_3 q_4, T T}
     + M^{\rho\sigma\mu\nu}_{q_1 q_2 q_3 q_4, T T} \bigr]
 &= \tilde{u}_{T T,A}^{\mu\nu\rho\sigma}\, A_{\delta (q_1 q_2) \delta (q_3 q_4)}^{}
    + \tilde{u}_{T T,B}^{\mu\nu\rho\sigma}\, m^2\ms B_{\delta (q_1 q_2) \delta (q_3 q_4)}^{}
\nonumber \\[0.3em]
 &\quad 	+ \tilde{u}_{T T,C}^{\mu\nu\rho\sigma}\, m^2\ms C_{\delta (q_1 q_2) \delta (q_3 q_4)}^{}
    + \tilde{u}_{T T,D}^{\mu\nu\rho\sigma}\, m^4\ms D_{\delta (q_1 q_2) \delta (q_3 q_4)}^{}
\nonumber \\[0.3em]
 &\quad   + u_{T T,E}^{\mu\nu\rho\sigma}\, m^2\ms \widetilde{E}_{\delta (q_1 q_2) \delta (q_3 q_4)}^{} \,,
\end{align}
The tensors $u$ and $\tilde{u}$ have been worked out in \cite{Bali:2021gel} and depend on the Lorentz vectors $y^\mu$ and $p^\mu$. The invariant functions $A_{a_1 a_2}$, $B_{a_1 a_2}$, etc are functions of the Lorentz scalars $\omega$ and $y^2$. At leading twist, only $A_{a_1 a_2}$ and $B_{\delta(q_1 q_2)\delta(q_3 q_4)}$ contribute. These \textit{twist-two functions} are connected to the first Mellin moment by a Fourier transform: 

\begin{align}
\label{eq:skewed-mellin-inv-fct}
I_{a_1 a_2}(\zeta, y^2)
&= \int_{-\infty}^{\infty} \dd\omega\, e^{-i\zeta \omega}\, A_{a_1 a_2}(\omega,y^2) \,,
\\
\label{eq:skewed-mellin-inv-fct-quad}
I^t_{\delta(q_1 q_2)\delta(q_3 q_4)}(\zeta, y^2)
&= \int_{-\infty}^{\infty} \dd\omega\, e^{-i\zeta \omega}\, B_{\delta(q_1 q_2)\delta(q_3 q_4)}(\omega,y^2) \,.
\end{align}
Note that \eqref{eq:skewed-mellin-inv-fct-quad} is only defined for transverse polarization of $q_1$ and $q_2$. 
%
%%%%%%%%%%%%%%%%%%%%%%%%%%%%%%%%%%%%%%%%
%

It has been found in \cite{Bali:2021gel, Reitinger:2024ulw, Diehl:2025kcr} that it is justified to assume that the DPD Mellin moments $I(\zeta,y^2)$ can be written in a factorized form w.r.t.\ their arguments, i.e.:

\begin{align}
   \label{eq:product-ansatz}
   I_{a_1 a_2}(\zeta,y^2)
   =
   J_{a_1 a_2}(\zeta) \,
   K_{a_1 a_2}(y^2)
   \,,
\end{align}

and likewise for $I^t_{\delta(q_1 q_2)\delta(q_3 q_4)}$. 
In this case, the function $K(y^2)$ is equivalent to respectively $A(\omega=0,y^2)$ or $B(\omega=0,y^2)$ up to a normalization factor, which is equal to one if we rewrite 

\begin{align}
    A(\omega,y^2) = \hat{A}(\omega) A(\omega=0,y^2)\,,
\end{align}
and similar for the twist-two function $B$.
This becomes obvious by inserting \eqref{eq:product-ansatz} in \eqref{eq:skewed-mellin-inv-fct} or \eqref{eq:skewed-mellin-inv-fct-quad}, respectively, and inverting the Fourier transform. Hence, if \eqref{eq:product-ansatz} holds, the twist-two functions $A_{a_1 a_2}$ and $B_{\delta(q_1 q_2)\delta(q_3 q_4)}$ fully describe the $y^2$ dependence of the DPD Mellin moments.

The determination of the function $J$, on the other hand, strongly relies on the knowledge of the dependence of the invariant functions on the Ioffe time $\omega$, since we have to invert the Fourier transform:

\begin{align}
   \label{eq:J-from-Ahat}
   \frac{1}{2\pi} \int_{-1}^{1} \dd \zeta\,
      e^{i \zeta \omega} \, J_{a_1 a_2}(\zeta)
   &=
   \hat{A}_{a_1 a_2}(\omega)
   \,,
\end{align}
It has been shown in \cite{Diehl:2025kcr} that both limitations on the lattice w.r.t.\ the Ioffe time $\omega$, as well as insufficient statistical precision, make an extraction of $J_{q_1 q_2}$ from the lattice data for $\hat{A}(\omega)$ unfeasible. Therefore, in this study, we focus on $K_{q_1 q_2}(y^2) = A(\omega = 0,y^2)$, which we can directly obtain from our lattice data for $\vec{p}=\vec{0}$.

\section{Lattice calculation details}
\label{sec:lattice-calc}

This section outlines the procedure used to evaluate the two-current matrix elements on the lattice by calculating four-point functions. A full treatment can be found in earlier publications~\cite{Bali:2021gel}.

%%%%%%%%%%%%%%%%%%%%%%%%%%%%%%%%%%%%%%%%

\subsection{Two-current matrix elements}

\begin{figure}
\begin{center}
\includegraphics[scale=1]{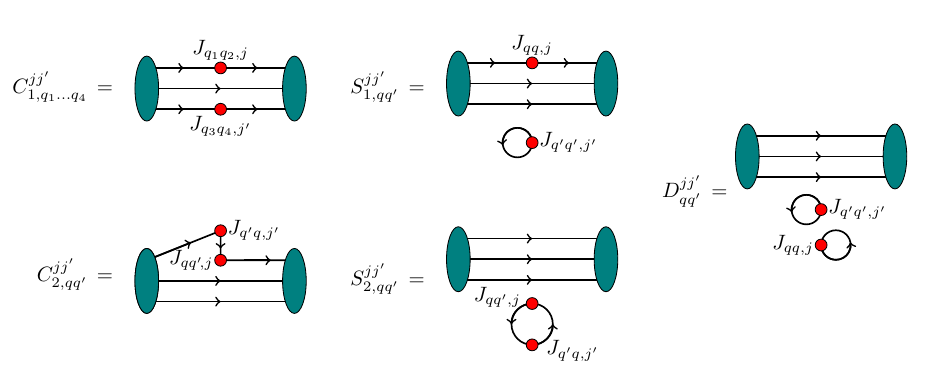}
\end{center}
\caption{Illustration of the five kinds of Wick contractions (graphs) contributing to a four-point function of a baryon. The explicit contributions for the graphs $C_1$, $C_2$ and $S_1$ depend on the quark flavor of the current insertions (red points). In the case where all quark flavors have the same mass, $C_2$ only depends on the flavors of the two propagators connected to the source or the sink. These flavors have to be the same for proton-proton matrix elements. \label{fig:graphs}}
\end{figure}

The two-current matrix element \eqref{eq:mat-els} is evaluated on the lattice after a Wick rotation to Euclidean space, where the time component is conventionally indexed by 4 rather than 0. To this end, we introduce the following four-point function for the proton:

\begin{align}
   C^{i j,\mvec{p}}_{\mathrm{4pt},\ms q q'}(\mvec{y},t,\tau)
   &=
   a^6 \, \sum_{\mvec{z}^\prime,\mvec{z}}
   e^{-i\mvec{p}(\mvec{z}^\prime-\mvec{z})}\
   \left\langle \tr \left\{
      P_+ \mathcal{P}(\mvec{z}^\prime,t)\ J_{q,i}(\mvec{y},\tau)\
      J_{q',j}(\mvec{0},\tau)\ \overline{\mathcal{P}}(\mvec{z},0)
   \right\} \right\rangle
   \,,
   \label{eq:4ptdef}
\end{align}
together with the two-point function

\begin{align}
   C^{\mvec{p}}_{\mathrm{2pt}}(t)
   &=
   a^6 \, \sum_{\mvec{z}^\prime,\mvec{z}}
   e^{-i\mvec{p}(\mvec{z}^\prime-\mvec{z})}\
   \left\langle \tr \left\{
      P_+ \mathcal{P}(\mvec{z}^\prime,t)\
      \overline{\mathcal{P}}(\mvec{z},0) \right\}
   \right\rangle
   \,.
   \label{eq:2ptdef}
\end{align}
The momentum projection is implemented through the phase factor $e^{-i\mvec{p}(\mvec{z}^\prime-\mvec{z})}$, which selects proton states carrying three-momentum $\mvec{p}$. The operator $J_{q, i}(\mvec{y},\tau)$ denotes the Euclidean counterpart of the Minkowski currents \eqref{eq:local-ops}, and $P_+$ is the positive-parity projector. The interpolating fields $\mathcal{P}(\mvec{z}^\prime,t)$ and $\overline{\mathcal{P}}(\mvec{z},0)$ respectively annihilate and create a baryon with the quantum numbers of the proton ($J=1/2$, $I=1/2$):

\begin{align}
   \overline{\mathcal{P}}(\mvec{x},t)
   &=
   \left.\epsilon_{a b c}\
      \left[
         \bar{u}_a(x)\ C \gamma_5\ \bar{d}_b^{\,T}(x)
      \right] \bar{u}_c(x)
   \right|_{x^4=t} \,,
   \nonumber \\[0.2em]
   \mathcal{P}(\mvec{x},t)
   &=
   \left.\epsilon_{a b c}\ u_a(x)
      \left[
         u_b^T(x)\ C \gamma_5\ d_c(x)
      \right]
   \right|_{x^4=t}
   \,,
\label{eq:interpdef}
\end{align}
where $C$ denotes the charge conjugation matrix in spinor space. The traces in \eqref{eq:4ptdef} and \eqref{eq:2ptdef} run over the open spinor indices of the quark fields $u_a$ and $\bar{u}_c$. We write $\tau$ for the time separation between the source and the current insertions. The ground-state matrix element \eqref{eq:mat-els} is recovered in the limit of large time separations:

\begin{align}
   \left. \vphantom{\sum}
   M_{q q', i j}(p,y)
   \right|_{y^0 = 0}
   &=
   2 V \sqrt{m^2 + \mvec{p}^2} \,
   \left.
      \frac{C^{i j,\mvec{p}}_{\mathrm{4pt},\ms q q'}(\mvec{y},t,\tau)}
      {C^{\mvec{p}}_\mathrm{2pt}(t)}
   \right|_{0 \ll \tau \ll t}
   \,.
\label{eq:4pt-2pt-ratio}
\end{align}
The four-point function \eqref{eq:4ptdef} receives contributions from multiple distinct Wick contractions of the quark fields, illustrated in Figure~\ref{fig:graphs}. The relevant combinations entering the physical matrix elements are:

\begin{align}
\left. M_{uudd, ij}(p,y)\right|_{y^0 = 0} 
&= 
	C^{ij,\mvec{p}}_{1,uudd}(\mvec{y}) + 
	S^{ij,\mvec{p}}_{1,u}(\mvec{y}) + 
	S^{ji,\mvec{p}}_{1,d}(-\mvec{y}) + 
	D^{ij,\mvec{p}}(\mvec{y})\,,
\nonumber \\
\left. M_{uuuu, ij}(p,y)\right|_{y^0 = 0} 
&= 
	C^{ij,\mvec{p}}_{1,uuuu}(\mvec{y}) + 
	C^{ij,\mvec{p}}_{2,u}(\mvec{y}) + 
	C^{ji,\mvec{p}}_{2,u}(-\mvec{y}) + 
	S^{ij,\mvec{p}}_{1,u}(\mvec{y}) + 
	S^{ji,\mvec{p}}_{1,u}(-\mvec{y})
\nonumber \\
&\quad + 
	S_{2}^{ij,\mvec{p}}(\mvec{y}) + 
	D^{ij,\mvec{p}}(\mvec{y})\,,
\nonumber \\
\left. M_{dddd, ij}(p,y)\right|_{y^0 = 0} 
&= 
	C^{ij,\mvec{p}}_{2,d}(\mvec{y}) + 
	C^{ji,\mvec{p}}_{2,d}(-\mvec{y}) + 
	S^{ij,\mvec{p}}_{1,d}(\mvec{y}) + 
	S^{ji,\mvec{p}}_{1,d}(-\mvec{y}) 
\nonumber \\
&\quad +
	S_{2}^{ij,\mvec{p}}(\mvec{y}) + 
	D^{ij,\mvec{p}}(\mvec{y})\,,
\nonumber \\
\left. M_{duud, ij}(p,y)\right|_{y^0 = 0} 
&= 
	C^{ij,\mvec{p}}_{1,duud}(\mvec{y}) + 
	C^{ij,\mvec{p}}_{2,d}(\mvec{y}) + 
	C^{ji,\mvec{p}}_{2,u}(-\mvec{y}) + 
	S^{ij,\mvec{p}}_{2}(\mvec{y})\,,
\nonumber \\
\left. M_{uddu, ij}(p,y)\right|_{y^0 = 0} 
&= 
	C^{ji,\mvec{p}}_{1,duud}(-\mvec{y}) + 
	C^{ij,\mvec{p}}_{2,u}(\mvec{y}) + 
	C^{ji,\mvec{p}}_{2,d}(-\mvec{y}) + 
	S^{ij,\mvec{p}}_{2}(\mvec{y})\,.
\label{eq:phys_me_decomp}
\end{align}
A systematic derivation of all matrix element decompositions and a detailed account of the Wick contraction procedure are provided in~\cite{Bali:2021gel}.

Renormalization of the bare lattice operator $J^{\mathrm{latt}}_{q,i}(y)$ proceeds multiplicatively via

\begin{align}
\label{eq:latt_op_ren}
J_{q,i}^{\overline{\mathrm{MS}}}(y) = Z_i\, J_{q,i}^{\mathrm{latt}}(y)\,,
\end{align}
where the factor $Z_i$ also converts to the $\overline{\mathrm{MS}}$ scheme. The renormalization constants $Z_V$ and $Z_A$ carry no anomalous dimension, while $Z_T$ depends on the renormalization scale. The numerical values used in this work, obtained in \cite{Bali_2021}, are presented in Table \ref{tab:renorm_a}.

\newpage

\subsection{Lattice setup}
\begin{figure}[h]
\includegraphics[trim={0 0 0 25}, clip, scale=0.7]{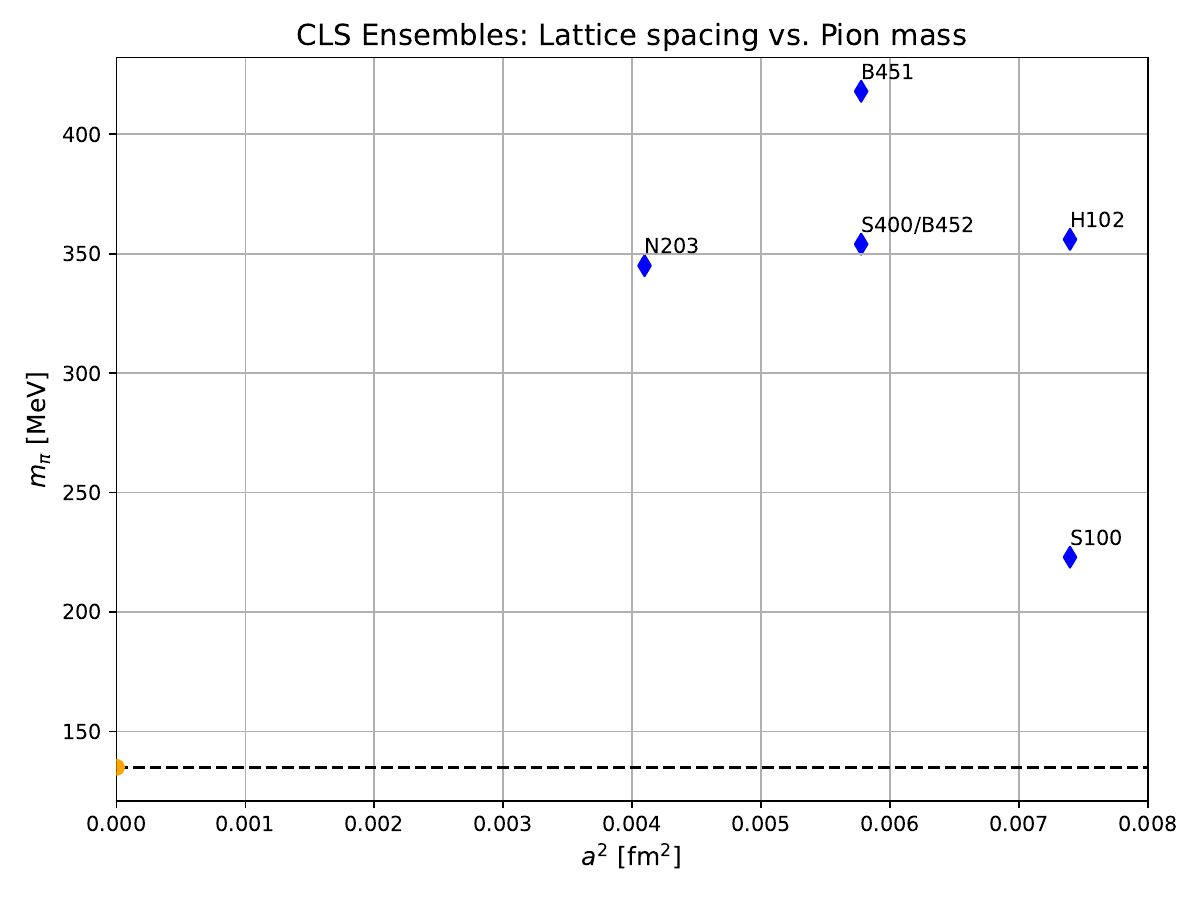}
\caption{Lattice spacing and pion masses of the CLS ensembles used in this work (blue diamonds) compared to the physical pion mass (dashed line) and the physical point (orange circle, bottom left). \label{fig:CLS-ens-overview}}
\end{figure}

Our goal is to investigate the physical limit of the lattice observables in the context of DPDs. To this end, we perform the simulations on several gauge ensembles, varying both the lattice spacing and light and strange quark masses. Moreover, the different physical volumes realized by the employed ensembles allow us to estimate finite volume effects. All simulations for this investigation are performed on ensembles provided by the CLS Collaboration \cite{Bruno_2015}. The ensembles used in this work have open boundary conditions in time direction except for B452 and B451, where antiperiodic boundary conditions are realized. The ensembles were generated using the tree level improved Lüscher-Weisz action and $n_f=2+1$ non-perturbatively $\mathcal{O}(a)$-improved Sheikholeslami-Wohlert fermions. A brief overview of the pion masses and lattice spacings of the ensembles used in this work is given in Figure \ref{fig:CLS-ens-overview}, details on all ensembles can be found in Table \ref{tab:cls}. 

\begin{table}[h]
\scriptsize
\begin{center}
\begin{tabular}{ccccccccc}
\hline
\hline
id & $\beta$ & $a[\mathrm{fm}]$  & $L^3 \times T$ & $\kappa_{l}$ & $\kappa_{s}$ & $m_{\pi,K}[\mathrm{MeV}]$ & $m_\pi L$ & $L$ [fm]  \\
\hline
H102 & $3.4$ & $0.085$ & $32^3 \times 96$  & $0.136865$ & $0.136549339$ & $354$, $442$ & $4.89$ & $2.72$  \\
S100 & $3.4$ & $0.085$ & $32^3 \times 128$ & $0.137030$ & $0.136222041$ & $214$, $476$ & $2.95$ & $2.72$  \\
S400 & $3.46$ & $0.075$ & $32^3 \times 128$ & $0.136984$ & $0.136702387$ & $354$, $445$ & $4.33$ & $2.41$  \\
B452 & $3.46$ & $0.075$ & $32^3 \times 64$  & $0.1370455$ & $0.136378044$ & $352$, $548$ & $4.31$ & $2.41$  \\
B451 & $3.46$ & $0.075$ & $32^3 \times 64$  & $0.136981435679729$ & $0.136408545268417$ & $422$, $577$ & $5.16$ & $2.41$  \\
N203 & $3.55$ & $0.064$ & $48^3 \times 128$ & $0.137080$ & $0.136840284$ & $348$, $445$ & $5.39$ & $3.06$  \\
\hline
\hline
\end{tabular}
\end{center}
\caption{Details of the CLS ensembles used in this work \cite{scale_setting}. \label{tab:cls}}
\end{table}

We use momentum smearing \cite{Bali:2016lva} with APE-smeared gauge links \cite{Albanese:1987} for both sources and sinks to improve the signal. APE smearing is iterated $n_\mathrm{APE}=25$ times with $\alpha_\mathrm{APE}=2.5$. Momentum smearing is performed using $b=0.45$ and $\epsilon=0.25$ at different iterations $n_\mathrm{mom}$, accounting for different lattice spacings and pion masses, see Table \ref{tab:momsmearparams}. The four-point functions are evaluated using the sequential source technique \cite{Martinelli:1988rr}, combining propagators from point sources as well as stochastic sources \cite{Foley:2005ac}. The stochastic propagators are improved by the hopping parameter expansion (HPE) \cite{Bali:2009hu} to reduce the noise introduced by them. The HPE is not used for the $C_1$ contractions, as their noise is not dominated by the stochastic propagator. The techniques used in the calculation of four-point functions as well as disconnected contributions are described in detail in \cite{Bali:2021gel}. The number of configurations and separation times used in this work is presented in Table~\ref{tab:ens_runs}. All simulations in this work have been performed using the Chroma software package \cite{Edwards:2004sx}.

\begin{table}[h!]
\begin{center}
\begin{tabular}{ccccc}
\hline
\hline
id & $|\vec{p}|~[\mathrm{GeV}]$ & $t_{\mathrm{sep}}[a]$ & $t_{\mathrm{sep}}[\mathrm{fm}]$ & \#cnfgs. used \\
\hline
H102 & $0$ & 12 & 1.020 & 990 \\
\hline
 &  & 8 & 0.68 & 890 \\
S100 & $0$ & 10 & 0.850 & 890 \\
 &  & 12 & 1.02 & 890 \\
\hline
 &  & 11 & 0.825 & 100 \\
S400 & $0$ & 13 & 0.975 & 1000 \\
 &  & 16 & 1.2 & 500 \\
\hline
B451/B452 & $0$ & 13 & 0.975 & 750 \\
\hline
N203 & $0$ & 16 & 1.024 & 300 \\
\hline
\hline
\end{tabular}
\end{center}
\caption{Summary of ensembles, momenta, source-sink separations and number of configurations used for the investigation of the lattice spacing and mass dependence. The results of the calculations on the H102 ensemble were already published in~\cite{Bali:2021gel} and~\cite{Reitinger:2024ulw}.}
\label{tab:ens_runs}
\end{table}

\begin{table}[h]
\begin{center}
\begin{tabular}{ccccccc}
\hline
\hline
 & H102 & S100 & S400 & B451 & B452 & N203 \\
\hline
$n_\mathrm{mom}$ & 250 & 350 & 310 & 270 & 310 & 445 \\
\hline
\hline
\end{tabular}
\end{center}
\caption{Number of momentum smearing iterations $n_\mathrm{mom}$ per CLS ensemble.\label{tab:momsmearparams}}
\end{table}

\begin{table}[h!]
\centering
\begin{tabular}{lccc}
\hline
\hline
 & $\beta = 3.40$ & $\beta = 3.46$ & $\beta = 3.55$ \\
\hline
$Z_V$ & 0.7115(5)(15) & 0.7206(2)(13) & 0.7327(3)(9) \\
$Z_A$ & 0.7510(4)(14) & 0.7580(3)(13) & 0.7683(4)(9) \\
$Z_T$ & 0.8294(4)(26) & 0.8429(2)(27) & 0.8619(3)(27) \\
\hline
\hline
\end{tabular}
\caption{Renormalization constants for the local currents on the ensembles used in the analysis at the renormalization scale $\mu = 2~\mathrm{GeV}$ \cite{Bali_2021} (table XXIX therein). For the analysis of the mass dependence, only the center column is required. \label{tab:renorm_a}}
\end{table}
The renormalization constants for all operators and ensembles analyzed in this work are shown in Table \ref{tab:renorm_a}, extracted from Table XXIX in \cite{Bali_2021}.

\newpage
\section{Results}
\label{sec:results}
The following analysis is organized as follows: First, we probe for excited state contributions within our data on one exemplary ensemble. Second, we examine the dependence of lattice artifacts observed in the past \cite{Bali:2021gel, Reitinger:2024ulw} on lattice parameters. Third, we present our results for the lattice spacing dependence of leading twist invariant functions, followed by their mass dependence as fourth point. The calculations on the H102 ensemble, the results of which will be used in the $a$-dependence analysis, have already been published in \cite{Bali:2021gel} and \cite{Reitinger:2024ulw}. In the following, we shall only discuss data from $C_1$ and $C_2$ contributions, as the disconnected contributions, $S_1$, $S_2$, and $D$, were found to be negligible compared to the connected ones and are therefore omitted at the present level of precision, in accordance with \cite{Bali:2021gel}.

\subsection{Excited state analysis}
Before extracting the matrix elements according to eq.~\eqref{eq:4pt-2pt-ratio}, we examine the plateau behavior of the correlators \eqref{eq:4ptdef} to assess potential excited-state contamination as well as signal strength and quality. Due to computation time limits, the only four-point contractions calculated for all intermediate insertion timeslices are the $C_1$ graphs, which we analyze in the following. 

\begin{table}[ht]
\begin{center}
\begin{tabular}{cccc}
\hline
\hline
EnsId & $t_{\mathrm{sep}}[a]$ & $t_{\mathrm{sep}}[\mathrm{fm}]$ & \#cnfgs. used \\
\hline
S400 &  11 & 0.836 & 100 \\
S400 &  13 & 0.988 & 1000 \\
S400 &  16 & 1.216 & 500 \\
\hline
S100 &  8 & 0.688 & 890 \\
S100 &  10 & 0.860 & 890 \\
S100 &  12 & 1.032 & 890 \\
\hline
\hline
\end{tabular}
\end{center}
\caption{Summary of separation times and number of configurations used for the excited state contribution analysis on S400 and S100 at momentum $\vec{p}=\vec{0}$.}
\label{tab:tsep_S400_S100}
\end{table}

To evaluate the validity and stability of the extraction, the calculations are repeated at different source-sink separations for the S400 and S100 ensembles, details given in Table~\ref{tab:tsep_S400_S100}.

\begin{figure}[ht]
\centering

%======================
% Row 1: V^0V^0, S100
%======================
\begin{subfigure}{0.49\textwidth}
\includegraphics[trim={0 0 0 28}, clip, width=\textwidth]{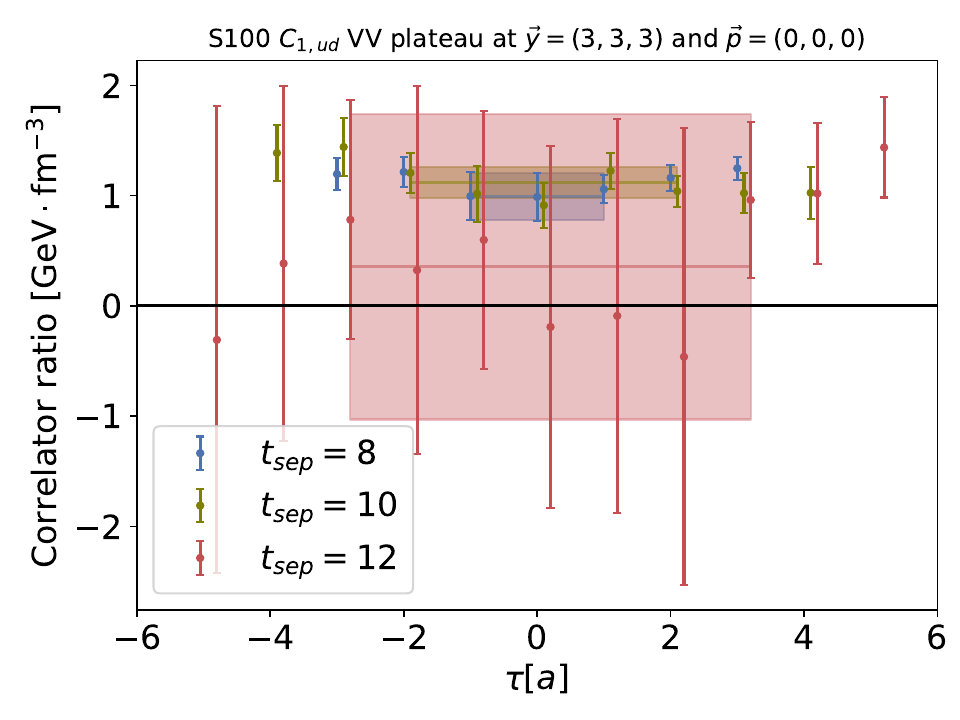}
\caption{$C_{1,ud}$, $\langle V^0V^0 \rangle$, S100}
\end{subfigure}
\hfill
\begin{subfigure}{0.49\textwidth}
\includegraphics[trim={0 0 0 28}, clip, width=\textwidth]{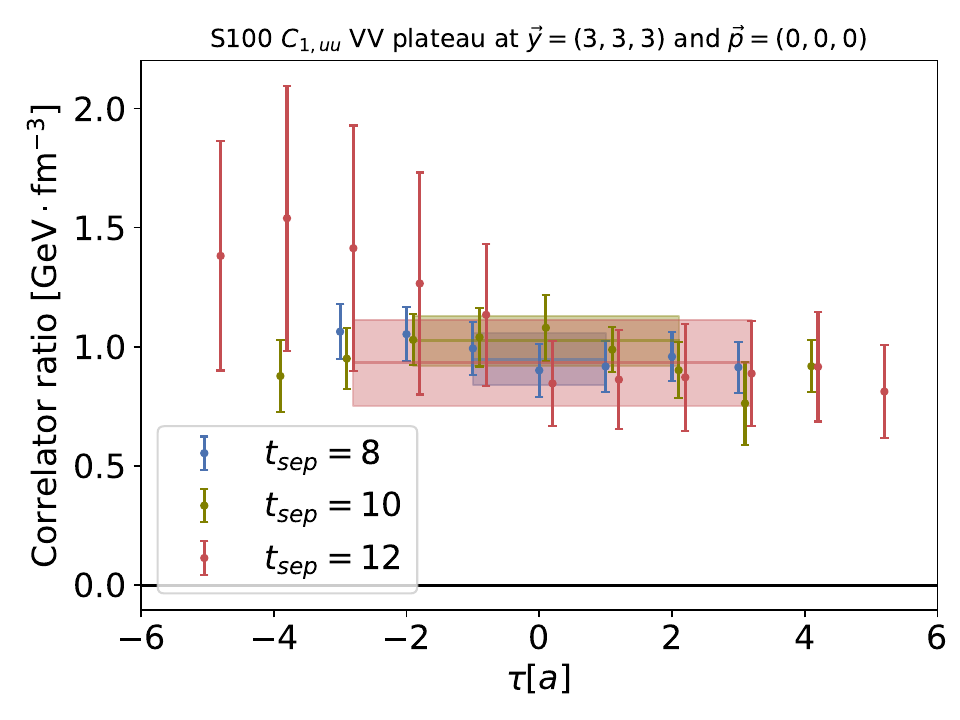}
\caption{$C_{1,uu}$, $\langle V^0V^0 \rangle$, S100}
\end{subfigure}

\vspace{0.5cm}

%======================
% Row 2: V^0V^0, S400
%======================
\begin{subfigure}{0.49\textwidth}
\includegraphics[trim={0 0 0 28}, clip, width=\textwidth]{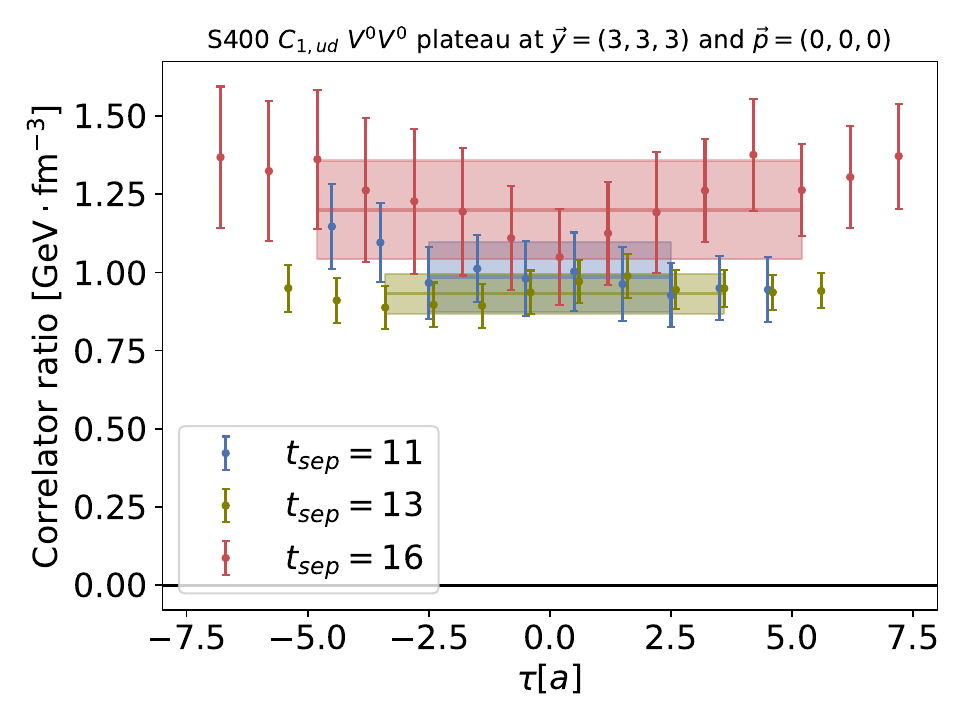}
\caption{$C_{1,ud}$, $\langle V^0V^0 \rangle$, S400}
\end{subfigure}
\hfill
\begin{subfigure}{0.49\textwidth}
\includegraphics[trim={0 0 0 28}, clip, width=\textwidth]{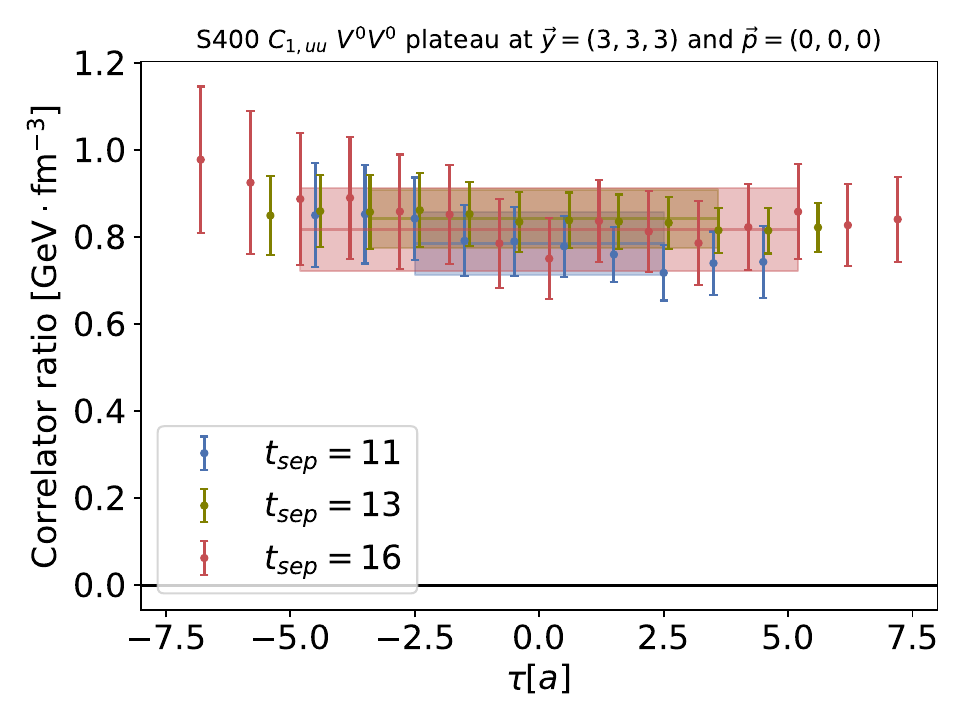}
\caption{$C_{1,uu}$, $\langle V^0V^0 \rangle$, S400}
\end{subfigure}

\caption{
Plateaus for the $C_{1,ud}$ (left column) and $C_{1,uu}$ (right column)
correlators on S100 (top row) and S400 (bottom row). All matrix elements are evaluated at $\vec{y}=(3,3,3)a$ and $\vec{p}=(0,0,0)$.
\label{fig:plateaus_S400_S100}}
\end{figure}

The results are shown in Figure~\ref{fig:plateaus_S400_S100}, where we show the ratios $R_{\mathrm{4pt}}$ to compare the results of different separation times. 
%The matrix element is calculated by plateau fit \eqref{eq:plateauextraction} along the insertion time $\tau$ using the fit range $\tau\in [3a,t-3a]$.
Only the data at $\vec{y}=(3,3,3)a$ with zero momentum are shown. We find that the plateaus obtained at different separations are consistent with each other within the statistical uncertainties. This implies that contamination by excited states is moderate. We thus extract the matrix elements by fitting a constant to the data around the central region in~$\tau$. The differences in statistical precision arise from both the varying separation times and, in the case of S400, the different numbers of configurations analyzed as listed in Table~\ref{tab:tsep_S400_S100}. We attribute the irregular shape present in the data with $t_{\mathrm{sep}}=11$ to the considerably decreased statistics, while the increasing errors at large $t_{\mathrm{sep}}$ are due to a decreasing SNR at larger timescales. In the following analysis, we focus on the data for $t=10a$ (S100) or $t=13a$ (S400), respectively, where we obtain a reasonable SNR, while excited state contamination is expected to be under control. For the ensembles B451, B452 and N203, all calculations are performed at only one separation time close to $1\,\mathrm{fm}$, details can be found in Table~\ref{tab:ens_runs}.

\begin{figure}[h]
\centering
\begin{subfigure}{0.49\textwidth}
\includegraphics[trim={0 0 0 28}, clip, width=\textwidth]{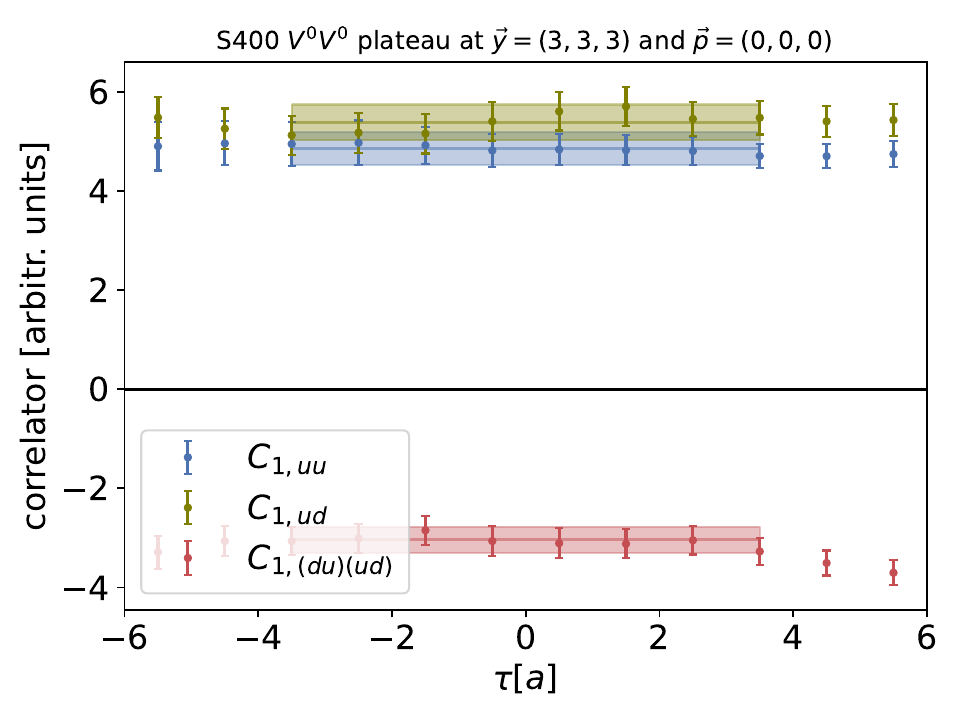}
\caption{S400, $\langle V^0V^0 \rangle$}
\end{subfigure}
\hfill
\vspace{0.5cm}
\begin{subfigure}{0.49\textwidth}
\includegraphics[trim={0 0 0 28}, clip, width=\textwidth]{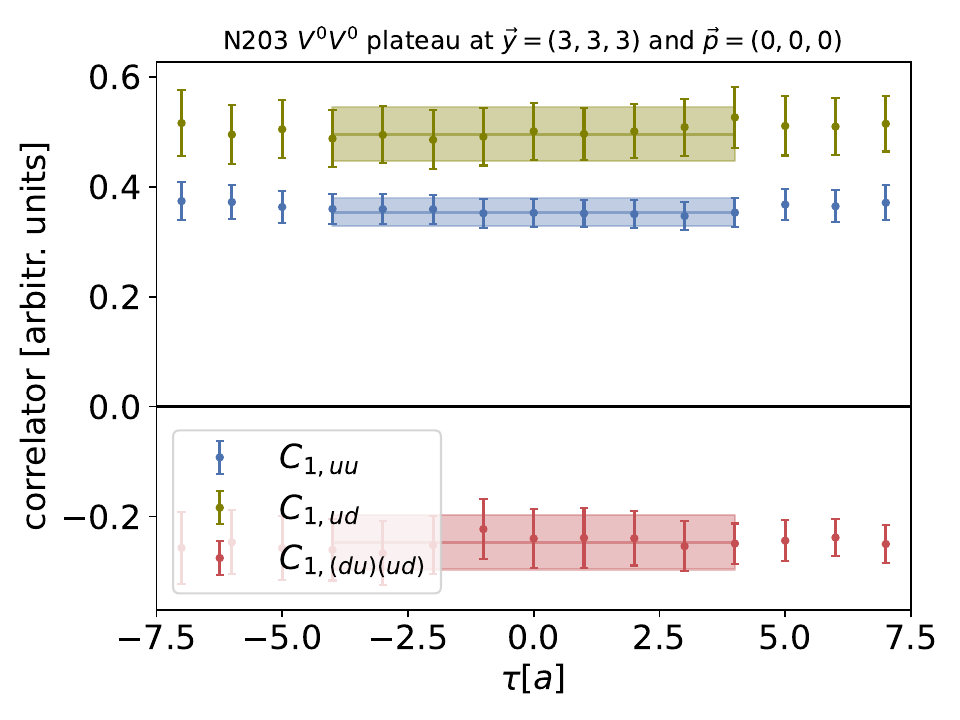}
\caption{N203, $\langle V^0V^0 \rangle$}
\end{subfigure}
\hfill
\begin{subfigure}{0.49\textwidth}
\includegraphics[trim={0 0 0 28}, clip, width=\textwidth]{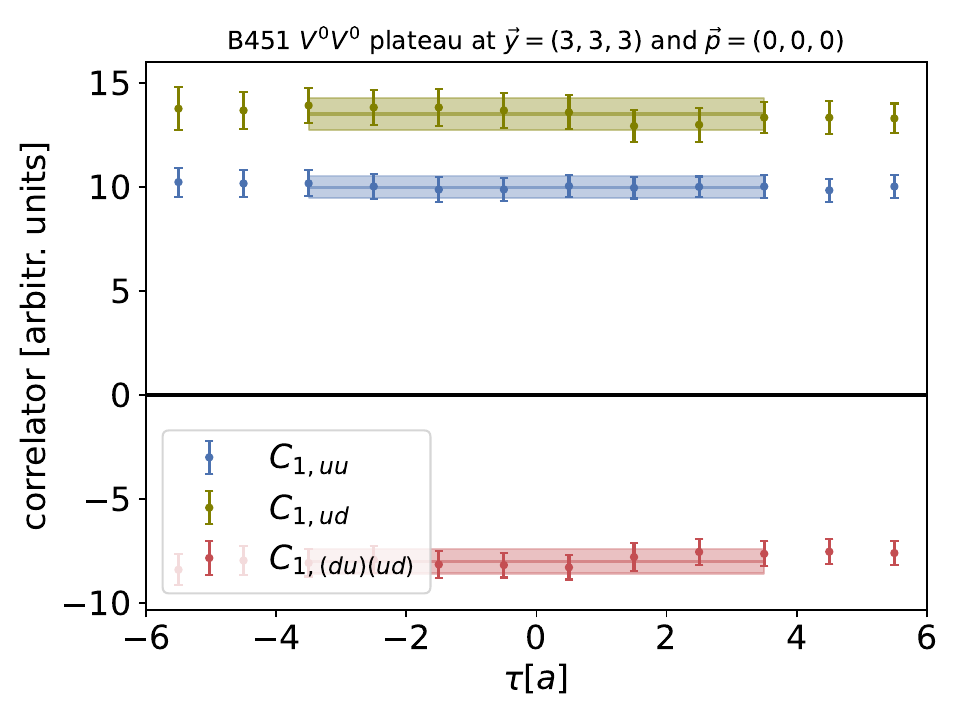}
\caption{B451, $\langle V^0V^0 \rangle$}
\end{subfigure}
\hfill
\begin{subfigure}{0.49\textwidth}
\includegraphics[trim={0 0 0 28}, clip, width=\textwidth]{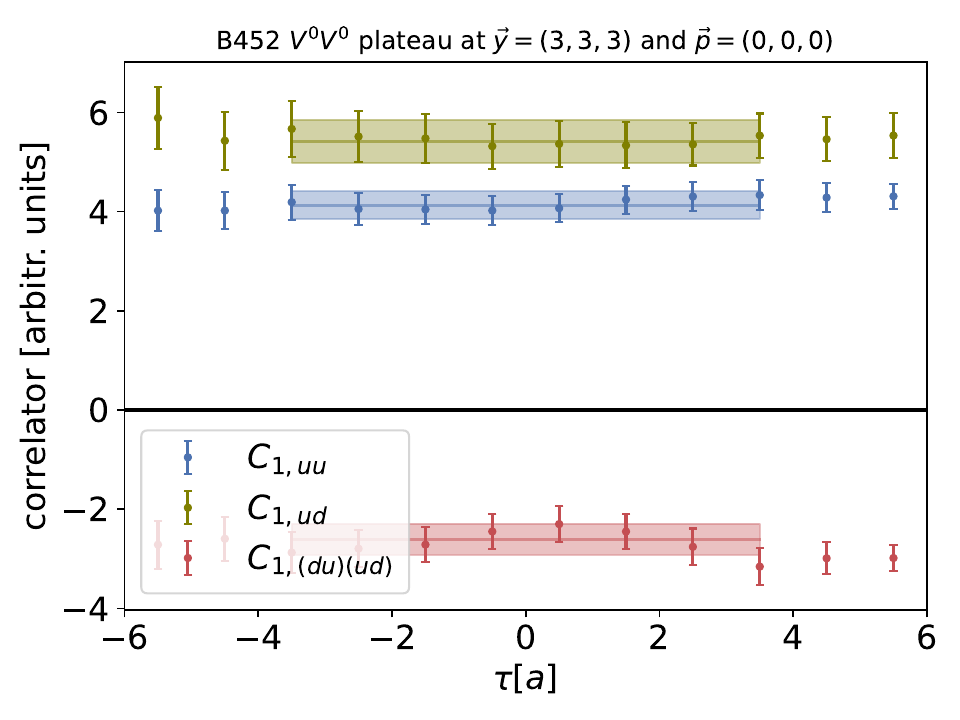}
\caption{B452, $\langle V^0V^0 \rangle$}
\end{subfigure}
\hfill
\vspace{0.2cm}
\begin{subfigure}{0.49\textwidth}
\includegraphics[trim={0 0 0 28}, clip, width=\textwidth]{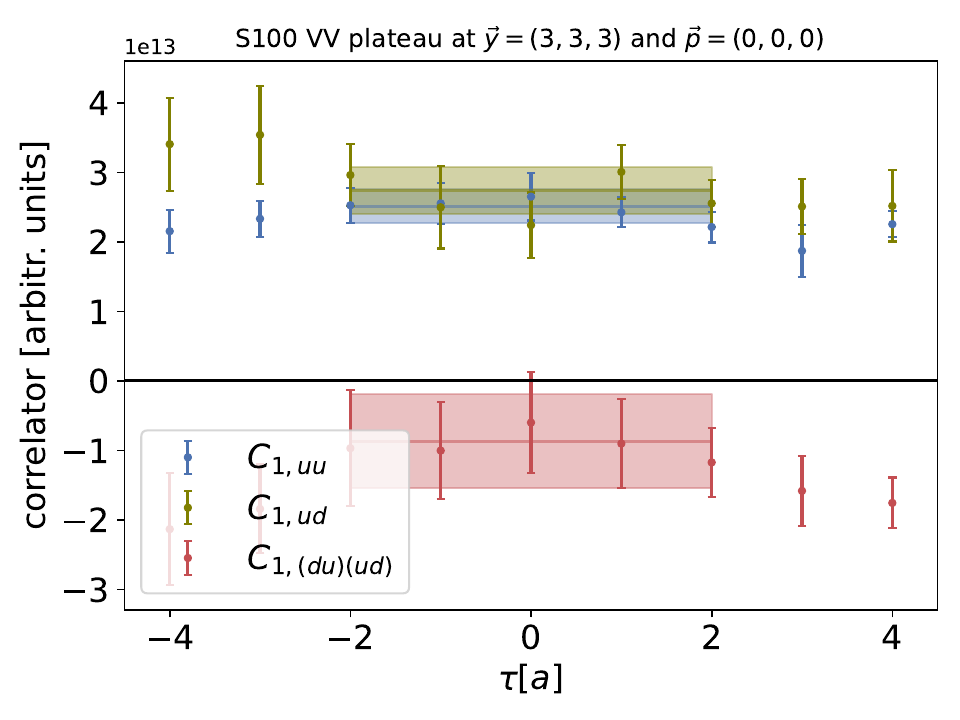}
\caption{S100, $\langle V^0V^0 \rangle$}
\end{subfigure}
\hfill
\caption{Comparison of the $C_1$ plateaus at $\vec{y}=(3,3,3)a$ for the $\langle V^0V^0 \rangle$  channel on the ensembles S400, N203, B451, B452 and S100 (from top left to bottom right). The best fit is presented as a straight line with errors, its extension indicating the subset of points used in the fit. \label{fig:ens_plat_comp}}
\end{figure}

Figure \ref{fig:ens_plat_comp} depicts the plateaus for the $C_1$ correlators for the $V^0V^0$ matrix element on the ensembles newly analyzed for this work. For the $V^0V^0$ channel, we observe plateaus extending over all insertion times. While the data for the other matrix elements is noisier than in the case of $V^0V^0$, we find no definite curvature in any of them either. We exclude points near the source and sink where residual excited-state effects may persist for added robustness of the extraction. 

\subsection{Lattice artifacts}
\label{sec:artifacts}
As the target quantities of our investigation are Lorentz invariant functions, they depend only on Lorentz invariants, i.e.\ $y^2$ and $p\cdot y=\omega$. This allows us to average the matrix element over all separations $\vec{y}$ of identical $(\omega,y^2)$\footnote{For ease of writing, we henceforth write $y$ for the distance $\sqrt{-y^2} =|\mvec{y}|$ when there is no risk of confusion.}, provided that symmetry breaking effects due to discretization are small. In order to have these effects under control, we investigate our data for artifacts stemming from both discretization and finite volume. 

\subsubsection{Finite volume effects}
\label{subsec:FV}
In order to investigate the potential impact of finite volume effects on the data, we analyze the dependence of the matrix elements and invariant functions on the direction of the separation $\vec{y}$. Specifically, we consider the angle $\theta$ enclosed between the vector $\vec{y}$ and the closest spatial diagonal $\hat{n}$, i.e.\ $\cos(\theta) = \vec{y}\cdot \hat{n}/|\vec{y}|$. Because of the periodic boundary conditions in the spatial direction, we expect the effects of mirror charges to be more pronounced along the lattice axis. This has already been done in Ref.~\cite{Bali:2021gel, Reitinger:2024ulw, Diehl:2025kcr}.

\begin{figure}[h]
\centering
\begin{subfigure}{0.49\textwidth}
\includegraphics[trim={0 0 0 0}, clip, width=\textwidth]{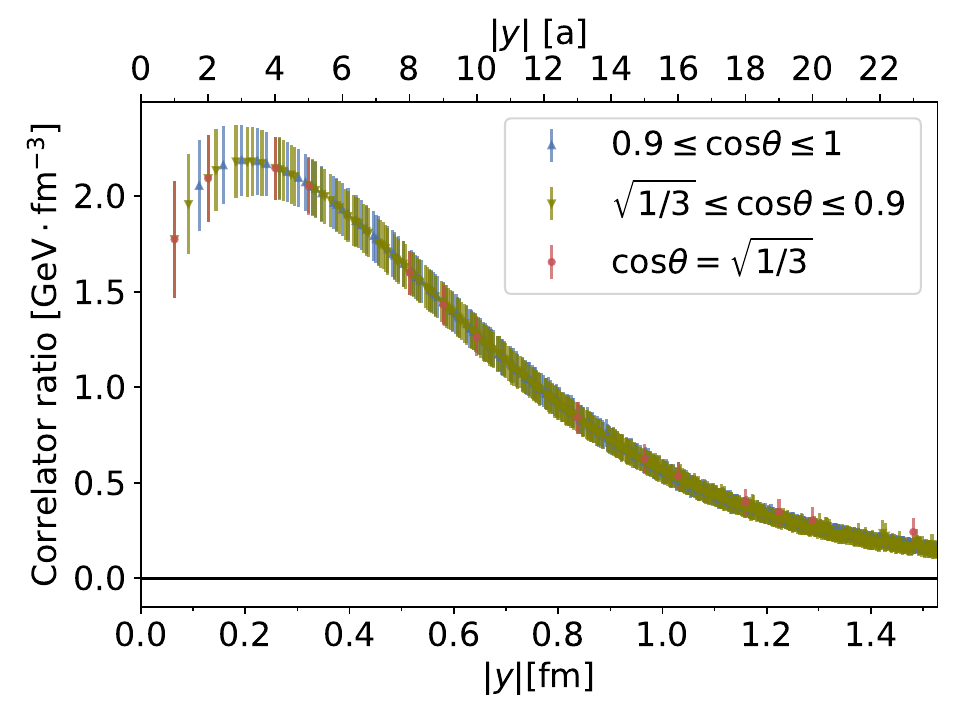}
\caption{$\langle V_u^0V_u^0 \rangle$ anisotropy for $C_{1,uuuu}$}
\end{subfigure}
\begin{subfigure}{0.49\textwidth}
\includegraphics[trim={0 0 0 0}, clip, width=\textwidth]{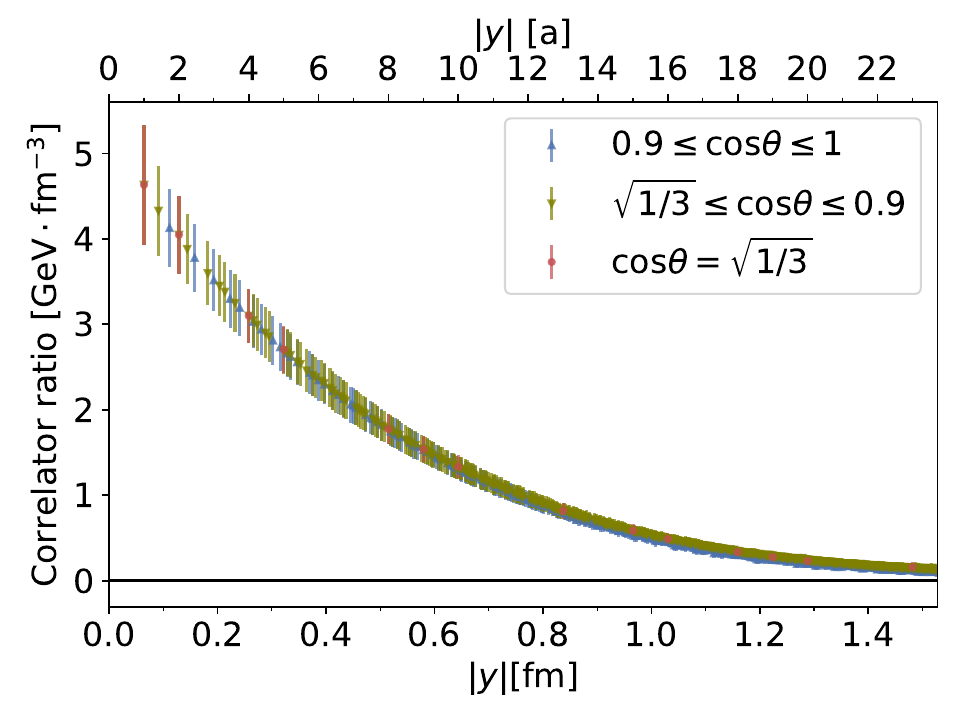}
\caption{$\langle V_u^0V_d^0 \rangle$ anisotropy for $C_{1,uudd}$}
\end{subfigure}
\begin{subfigure}{0.49\textwidth}
\includegraphics[trim={0 0 0 0}, clip, width=\textwidth]{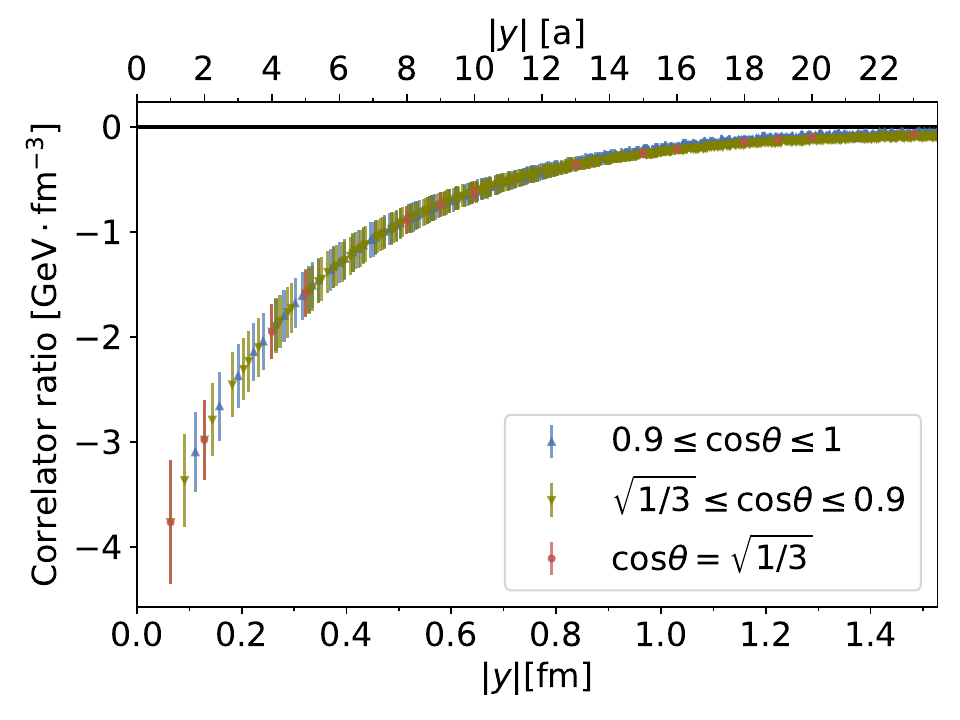}
\caption{$\langle V_{du}^0V_{ud}^0 \rangle$ anisotropy for $C_{1,duud}$}
\end{subfigure}
\caption{Anisotropy for $\langle V^0V^0 \rangle$ matrix elements from $C_1$ contractions on the N203 ensemble.  \label{fig:Aniso_N203}}
\end{figure}

A set of rotationally invariant matrix elements obtained from the N203 ensemble is shown in Figure~\ref{fig:Aniso_N203}. In the case of N203, the $C_1$ contraction exhibits no significant anisotropy effects at large $|y|$. This is in contrast to the case of H102, where a saw tooth pattern is clearly observable in  Figure 7 in Ref.~\cite{Bali:2021gel}. In all cases, including H102 \cite{Bali:2021gel, Reitinger:2024ulw}, the $C_2$ contraction has a pronounced anisotropy at smaller $|\vec{y}|$, independent of lattice spacing and extension.

\begin{figure}[h]
\centering
\begin{subfigure}{0.49\textwidth}
\includegraphics[trim={0 0 0 0}, clip, width=\textwidth]{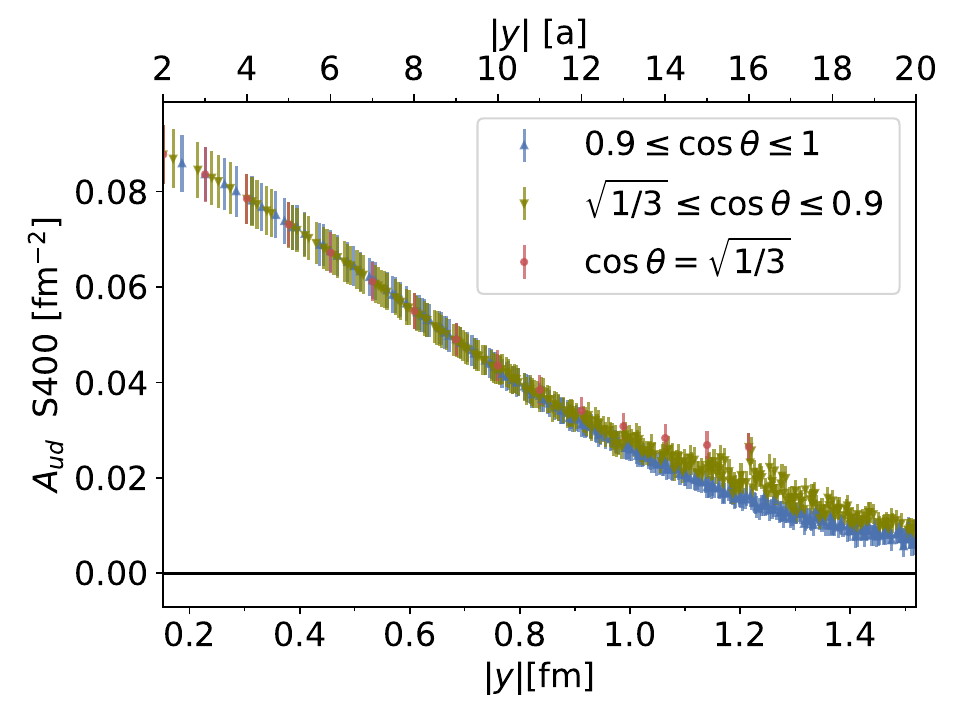}
\caption{S400, $A_{ud}$}
\end{subfigure}
\begin{subfigure}{0.49\textwidth}
\includegraphics[trim={0 0 0 0}, clip, width=\textwidth]{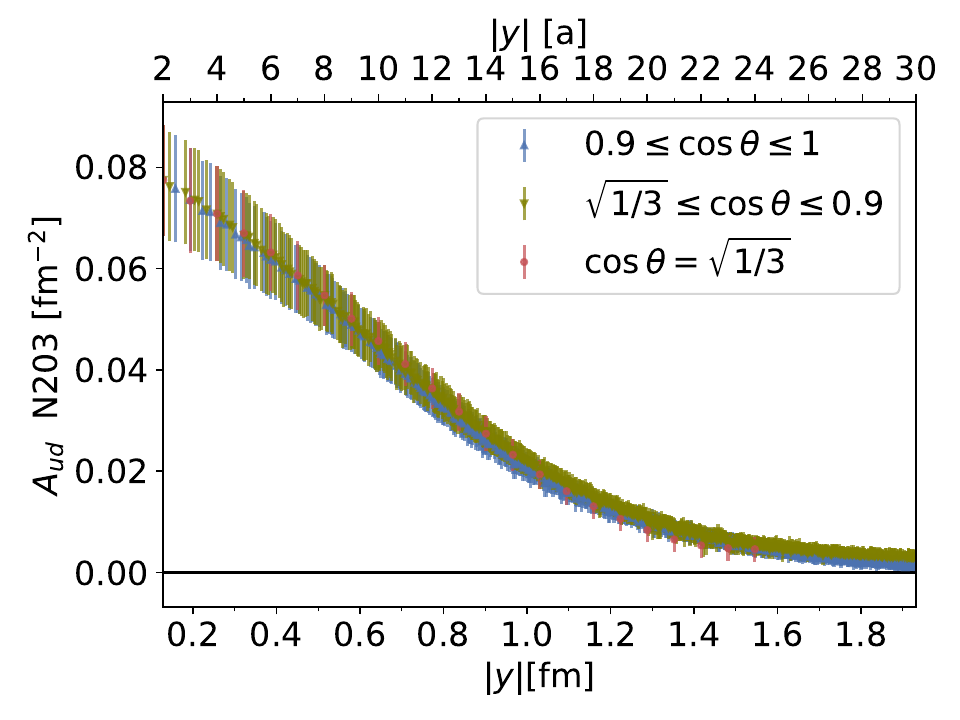}
\caption{N203, $A_{ud}$}
\end{subfigure}
\begin{subfigure}{0.49\textwidth}
\includegraphics[trim={0 0 0 0}, clip, width=\textwidth]{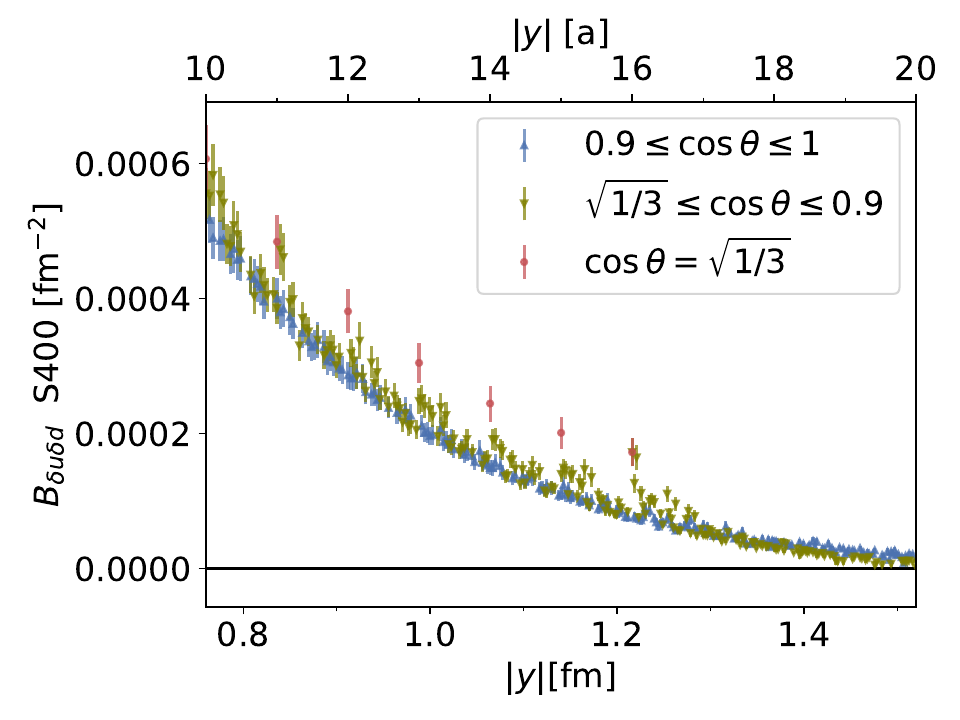}
\caption{S400, $B_{\delta u \delta d}$}
\end{subfigure}
\begin{subfigure}{0.49\textwidth}
\includegraphics[trim={0 0 0 0}, clip, width=\textwidth]{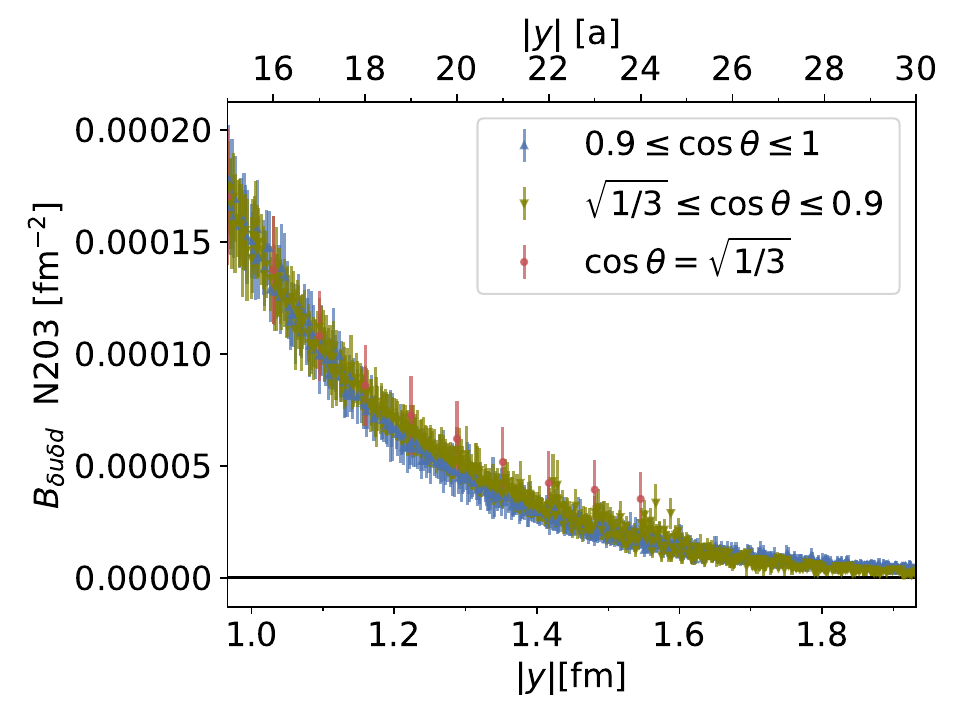}
\caption{N203, $B_{\delta u \delta d}$ \label{subfig:Aniso_a_dep_C1_N203_B}}
\end{subfigure}
\caption{Anisotropy comparison of $ud$ leading twist invariant functions $A$ and $B$ at $\vec{p}=\vec{0}$, only considering $C_1$ contraction contributions calculated on S400 (left column) and N203 (right column). $A$ for the vector-vector channel is shown in the top row, $B$ for the tensor-tensor channel is depicted in the bottom row. The included $|\vec{y}|\,\mathrm{[a]}$ range has been adapted to the spatial geometry of the ensembles. \label{fig:Aniso_a_dep_C1}}
\end{figure}

Figure \ref{fig:Aniso_a_dep_C1} shows selected leading twist invariant functions of the $ud$ flavor combination, with the Lorentz decomposition being performed solely on data from the $C_1$ contraction at hadron momentum $\vec{p}=\vec{0}$. The anisotropy effects observed on the S400 ensemble are very similar to those found in Ref.~\cite{Bali:2021gel} for H102 in all channels, exhibiting a clear saw tooth pattern at large separations $|\vec{y}|$. The ensembles S400 and H102 have smaller spatial extent and finite volume parameter than N203, see Table \ref{tab:cls}. For the N203 ensemble with $Lm_\pi=5.39$ and $La=3.06\,\mathrm{fm}$, finite volume effects are strongly suppressed, particularly in the vector-vector channel (first row). This can be explained by the increased distance between mirror images introduced by the periodic boundary conditions in space. Despite this, even on N203 residual finite volume effects remain, see e.g.\ Figure \ref{subfig:Aniso_a_dep_C1_N203_B}, necessitating an exclusion of all points with $\cos{\theta}<0.9$.

\begin{figure}[h]
\centering
\begin{subfigure}{0.49\textwidth}
\includegraphics[trim={0 0 0 0}, clip, width=\textwidth]{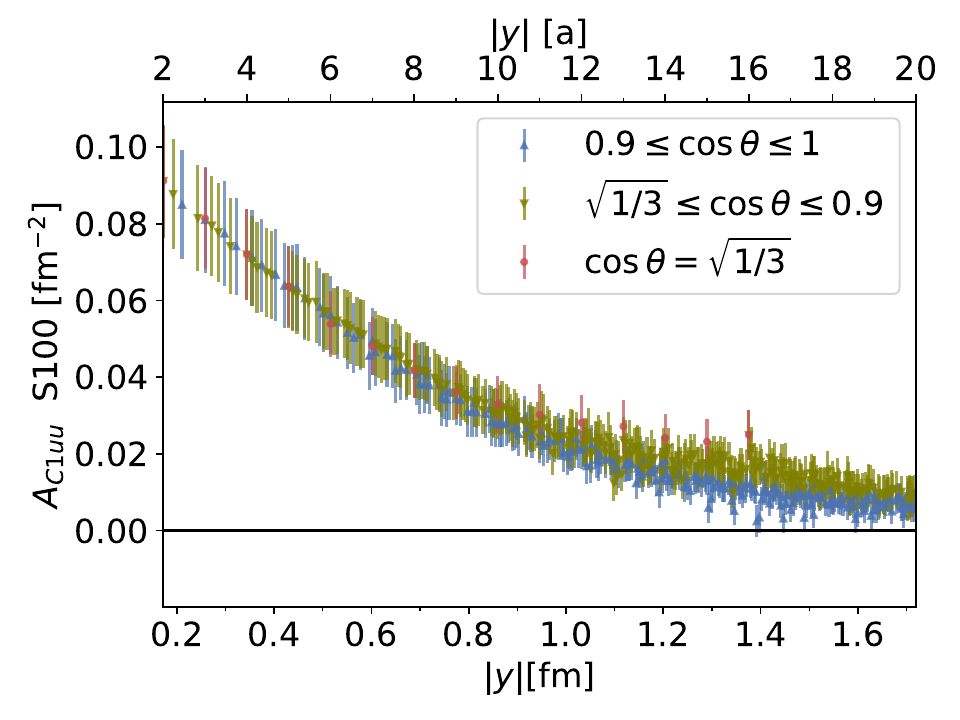}
\caption{S100, $Lm_\pi=2.95$}
\end{subfigure}
\begin{subfigure}{0.49\textwidth}
\includegraphics[trim={0 0 0 0}, clip, width=\textwidth]{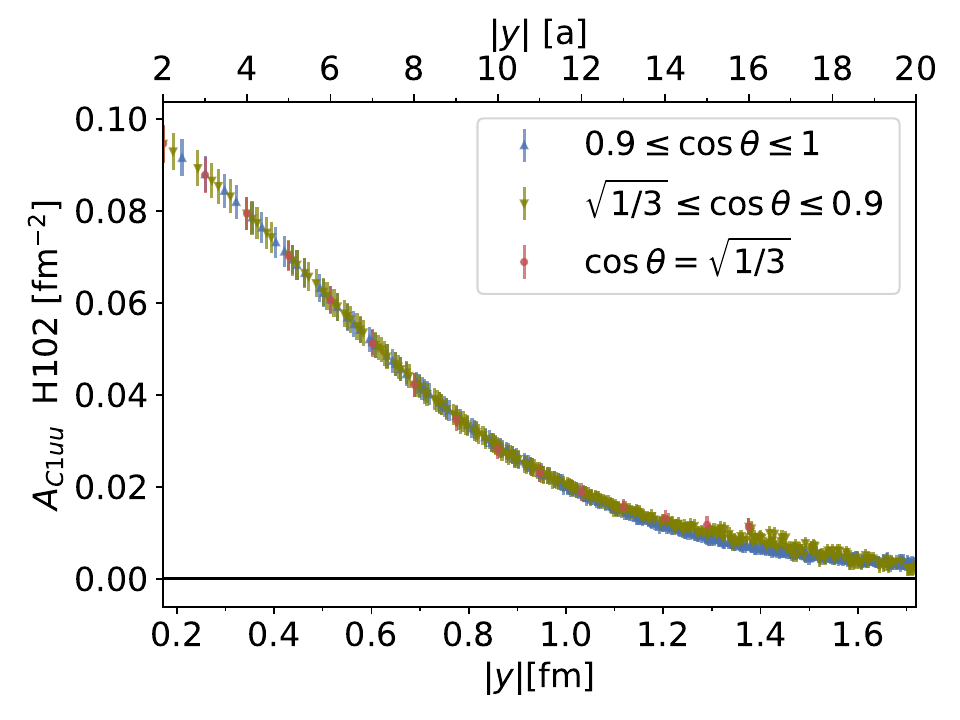}
\caption{H102, $Lm_\pi=4.89$}
\end{subfigure}
\caption{Comparison of invariant functions $A_{uu}$, considering only the $C_1$ contraction, on the S100 (left) and H102 (right) with separations $\vec{y}$ enclosing different angles $\theta$ with the diagonals. \label{fig:FV_S100_H102}}
\end{figure}

The same saw-tooth anisotropy pattern is also observed when comparing data obtained on the S100 and H102 ensembles, presented in Figure \ref{fig:FV_S100_H102}. Both ensembles have the same lattice spacing and spatial extension $La$, but the pion mass on S100 ($m_\pi=214\,\mathrm{MeV}$) is considerably smaller than on H102 ($m_\pi=354\,\mathrm{MeV}$), leading to their considerably different values of $Lm_\pi$.

\begin{figure}[h]
\centering
\begin{subfigure}{0.49\textwidth}
\includegraphics[trim={0 0 0 0}, clip, width=\textwidth]{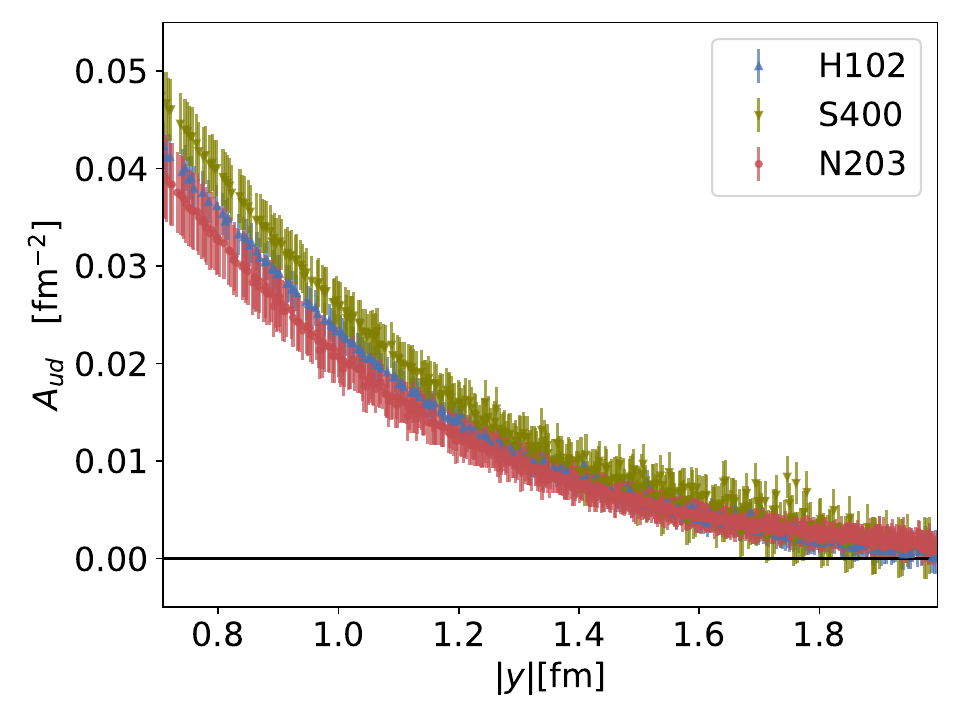}
\caption{$A_{ud}$}
\end{subfigure}
\begin{subfigure}{0.49\textwidth}
\includegraphics[trim={0 0 0 0}, clip, width=\textwidth]{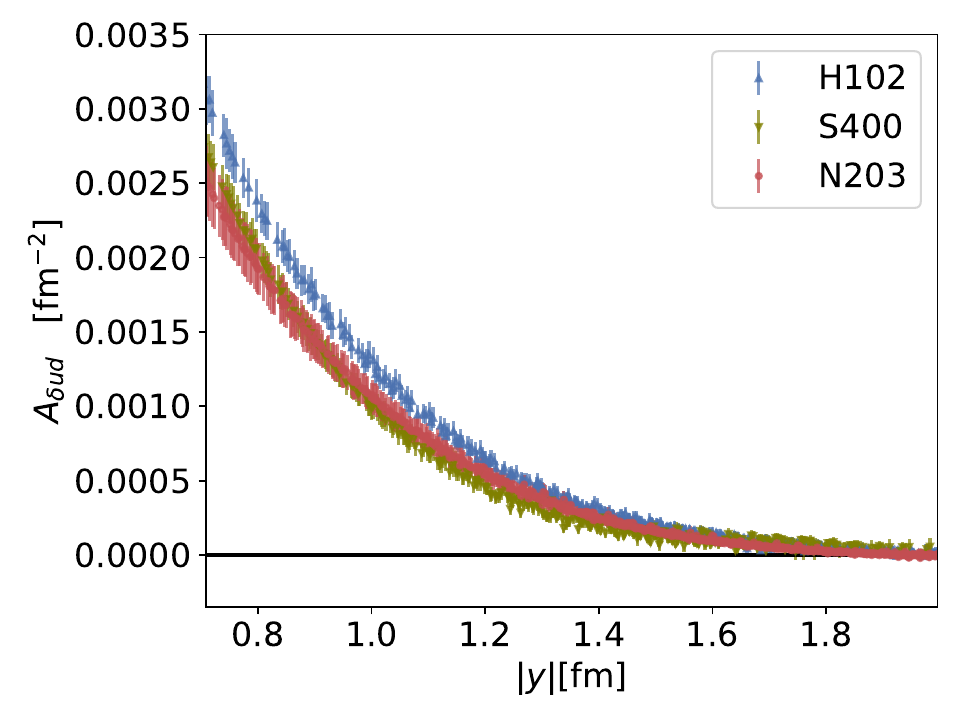}
\caption{$A_{\delta ud}$}
\end{subfigure}

\caption{Comparison of invariant functions $A_{ud}$ in the vector-vector channel (left) and tensor-vector channel (right) including separations $\vec{y}$ close to the lattice diagonals, such that $\cos{\theta}>0.9$ with extended range towards the boundary, obtained by computations on the H102 \cite{Bali:2021gel}, S400 and N203 ensembles. \label{fig:boundary_aniso}}
\end{figure}

It remains the question to what extent finite volume effects remain after the cut in $\cos\theta$. Figure \ref{fig:boundary_aniso} compares the results for three ensembles with similar $m_\pi\approx350\,\mathrm{MeV}$ but different physical lattice volumes ($2.41~\mathrm{fm}\le La \le 3.06~\mathrm{fm}$). We find that the data points agree within errors for large distances $\vec{y}$, indicating that finite volume effects at large separations are sufficiently under control after the cut in $\cos{\theta}$.

\subsubsection{Discretization artifacts}
\label{subsec:Disc}

\begin{figure}[h]
\centering
\begin{subfigure}{0.49\textwidth}
\includegraphics[trim={0 0 0 0}, clip, width=\textwidth]{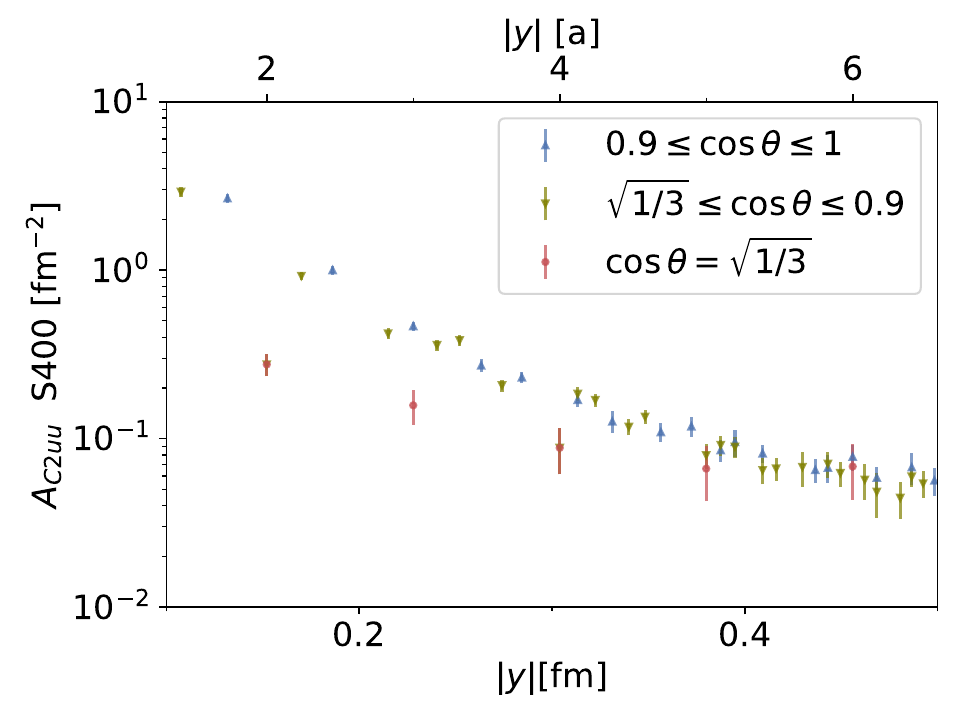}
\caption{S400, $A_{uu}$}
\end{subfigure}
\begin{subfigure}{0.49\textwidth}
\includegraphics[trim={0 0 0 0}, clip, width=\textwidth]{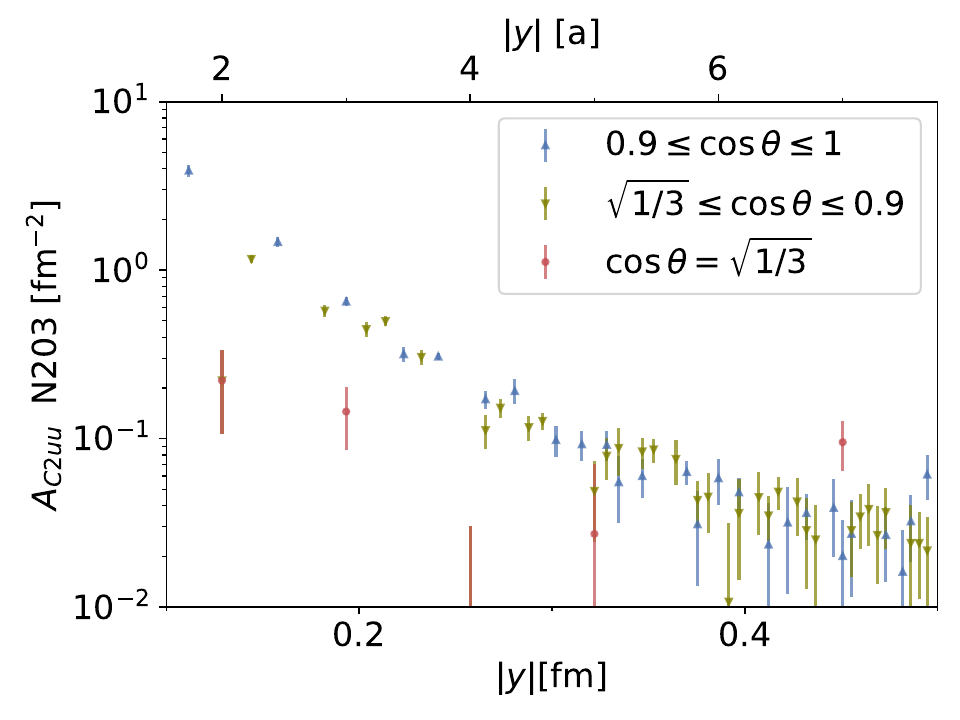}
\caption{N203, $A_{uu}$}
\end{subfigure}
\begin{subfigure}{0.49\textwidth}
\includegraphics[trim={0 0 0 0}, clip, width=\textwidth]{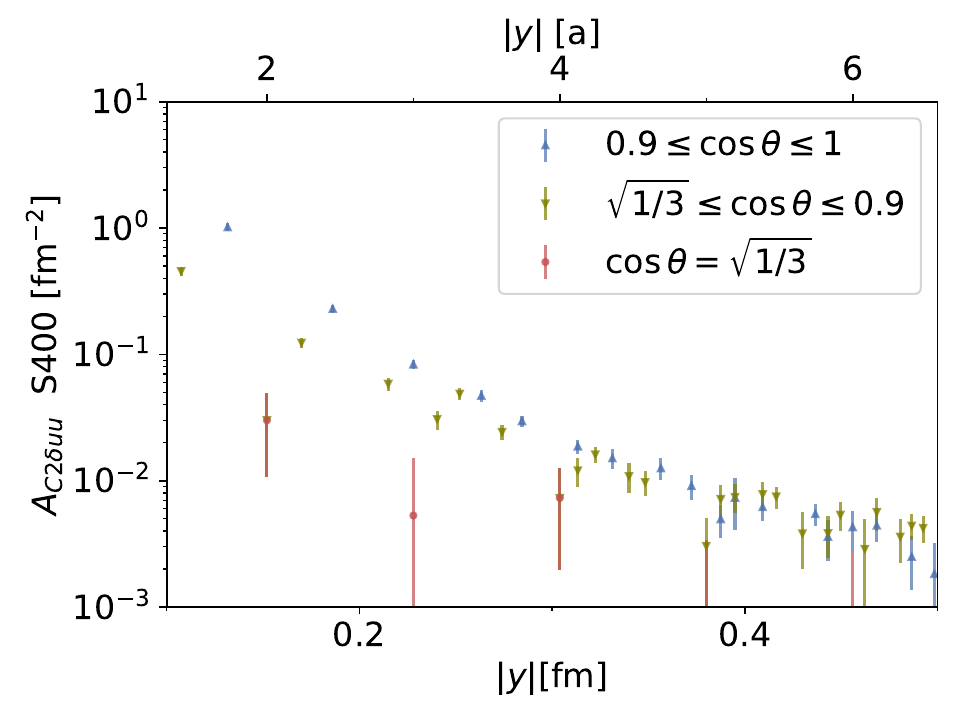}
\caption{S400, $A_{\delta u  u}$}
\end{subfigure}
\begin{subfigure}{0.49\textwidth}
\includegraphics[trim={0 0 0 0}, clip, width=\textwidth]{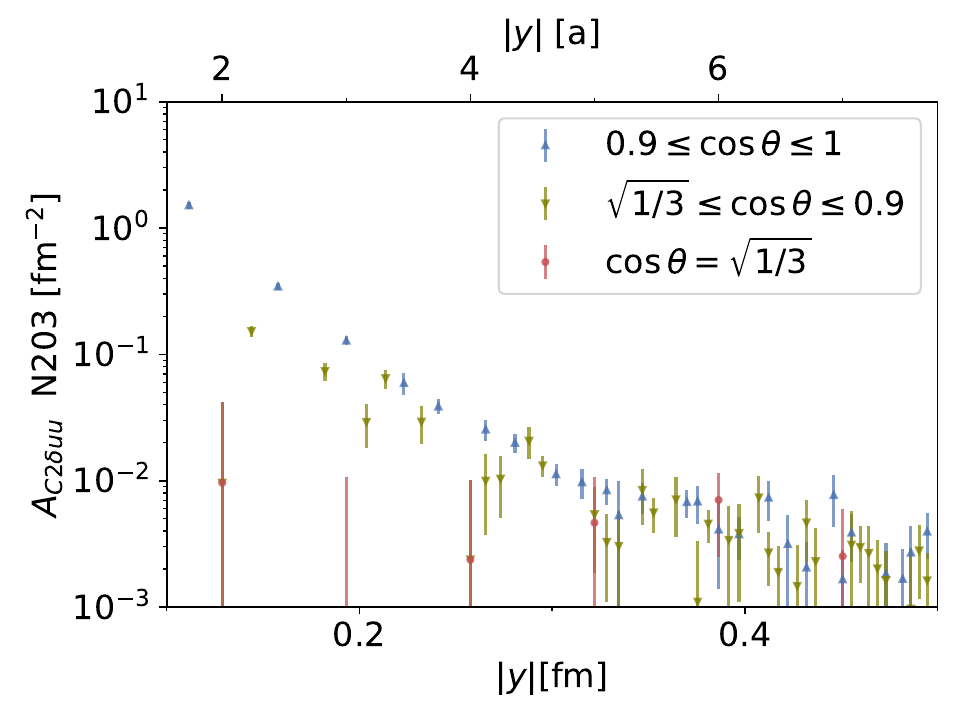}
\caption{N203, $A_{\delta u  u}$}
\end{subfigure}
\caption{Anisotropy comparison of $uu$ leading twist invariant functions $A$ at $\vec{p}=\vec{0}$, only considering $C_2$ contraction contributions calculated on S400 (left column) and N203 (right column). $A$ for the vector-vector channel is shown in the top row, the tensor-tensor channel is depicted in the bottom row. The limits of the physical distance $|y|\,[\mathrm{fm}]$ have been kept constant across all four subfigures. \label{fig:Aniso_a_dep_C2}}
\end{figure}

We proceed in similar fashion to \ref{subsec:FV} for finite volume effects. The same selection of invariant functions presented in Figure~\ref{fig:Aniso_a_dep_C1} is shown here again for the $uu$ flavor combination, only taking into account data from the $C_2$ contraction in Figure \ref{fig:Aniso_a_dep_C2}. As can be observed there, the data shows a strong anisotropic behavior at very small distances. This can be explained by discretization artifacts in the Dirac propagator for Wilson fermions, which have been discussed in, e.g.\ \cite{Bali:2017gfr, Bali:2018spj}. 

Notice that the anisotropy pattern in Figure \ref{fig:Aniso_a_dep_C2} is similar on both sides, with the exception that the points in panel (b) are shifted to the left compared to panel (a). This is a consequence of the smaller lattice spacing $a = 0.064~\mathrm{fm}$ in (b) compared to $a = 0.075~\mathrm{fm}$ in (a). The anisotropy artifacts become significant once $|y| < 5a$ for all considered lattice spacings. Hence, it could be expected that, for a given current distance in physical units, discretization artifacts will decrease for smaller lattice spacing.

\subsection{Lattice spacing dependence}
\label{sec:adep}

In this section we investigate the lattice spacing dependence of the leading twist invariant functions. We adhere to the data selection criteria established in Section \ref{sec:artifacts}, namely that the angle $\theta$ enclosed by the separations $\vec{y}$ with the closest lattice diagonal has to fulfill $\cos{\theta}>0.9$ and $4a\leq|\vec{y}|\leq La/2$. We thus account for discretization effects and finite volume artifacts discussed in the previous section. We will first discuss our findings for single contractions before considering the overall contribution for specific flavor combinations.

\subsubsection{$C_1$ contributions}
\label{subsec:a_dep_C1}

\begin{figure}[ht]
\centering

%================================================
% Row 1: VV
%================================================
\begin{subfigure}{0.49\textwidth}
\includegraphics[width=\textwidth]{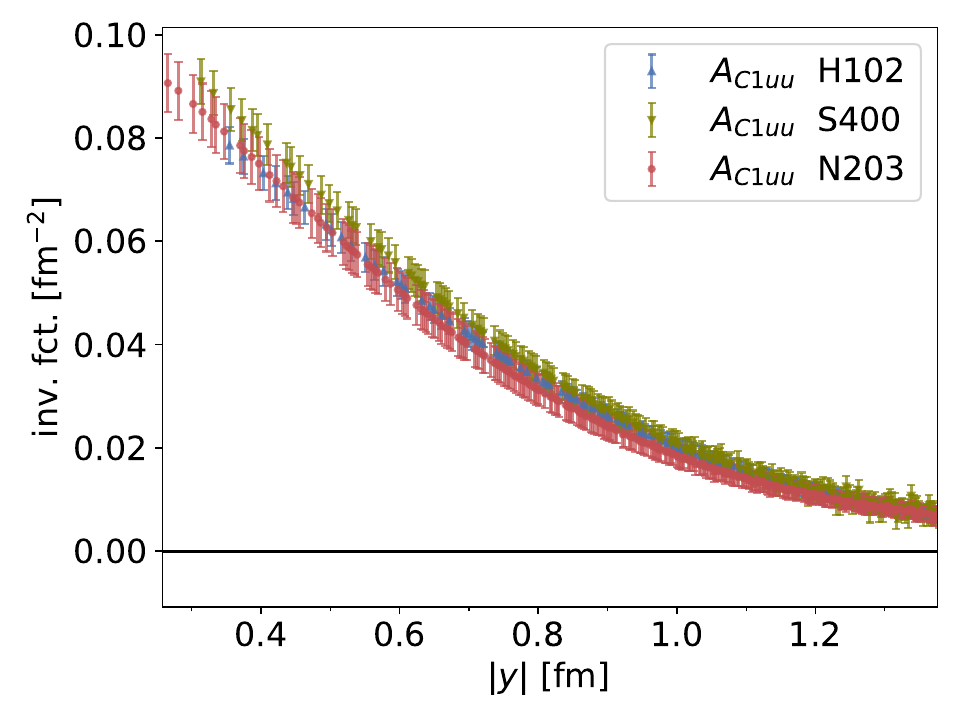}
\caption{$\mathrm{VV}$, $C_{1,uuuu}$}
\label{fig:inv_a_dep_VV_uuuu}
\end{subfigure}
\hfill
\begin{subfigure}{0.49\textwidth}
\includegraphics[width=\textwidth]{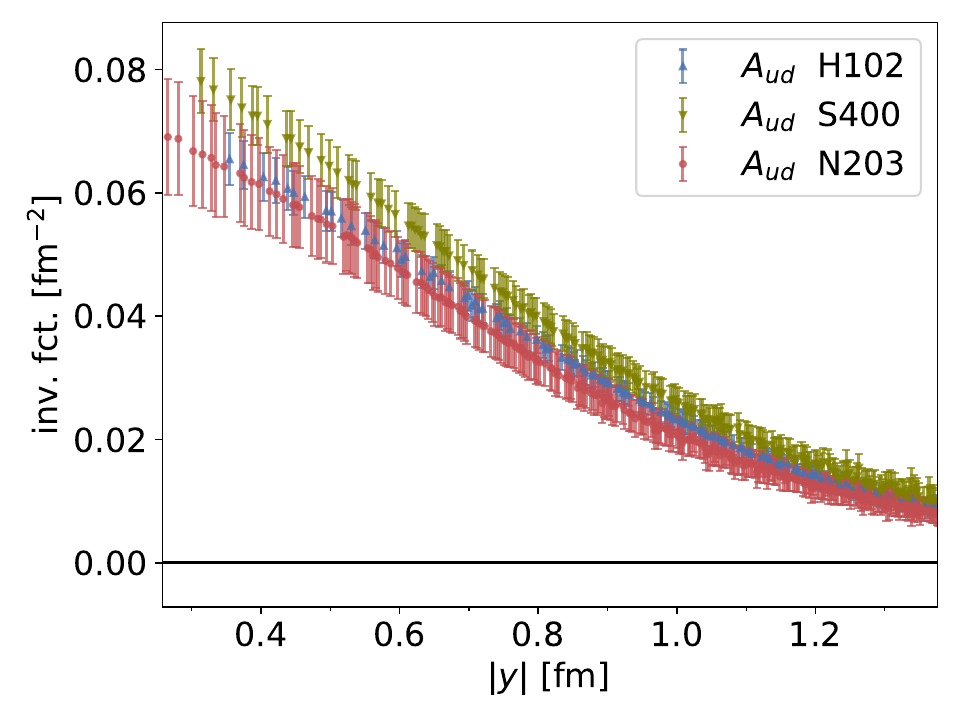}
\caption{$\mathrm{VV}$, $C_{1,ud}$}
\label{fig:inv_a_dep_VV_ud}
\end{subfigure}

%================================================
% Row 2: TV
%================================================
\begin{subfigure}{0.49\textwidth}
\includegraphics[width=\textwidth]{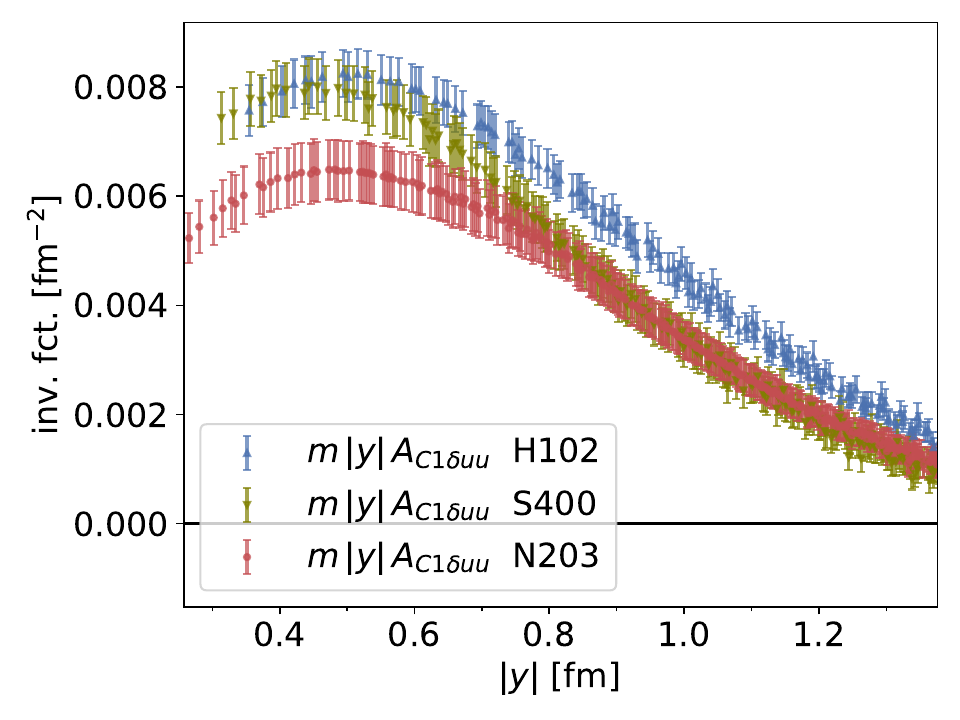}
\caption{$\mathrm{TV}$, $C_{1,uuuu}$}
\label{fig:inv_a_dep_TV_uuuu}
\end{subfigure}
\hfill
\begin{subfigure}{0.49\textwidth}
\includegraphics[width=\textwidth]{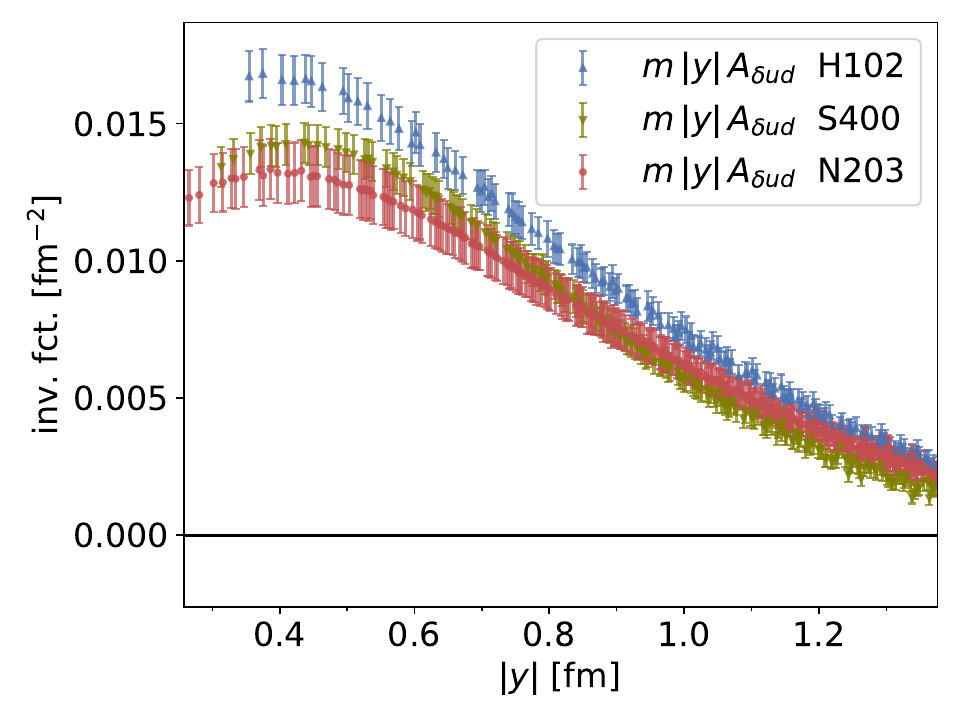}
\caption{$\mathrm{TV}$, $C_{1,ud}$}
\label{fig:inv_a_dep_TV_ud}
\end{subfigure}

\caption{
Leading twist invariant functions for selected channels. We plot the vector-vector (first row) and tensor-vector (second row) channels for the flavor combinations $C_{1,uuuu}$ (left column) and $C_{1,ud}$ (right column). Results are shown for the
H102 ($a=0.085\,\mathrm{fm}$), S400 ($a=0.075\,\mathrm{fm}$), and N203 ($a=0.064\,\mathrm{fm}$) ensembles at $\vec{p}=(0,0,0)$ and $\cos\theta>0.9$, only including data from the $C_1$ contraction.
}
\label{fig:inv_a_dep_all_channels}
\end{figure}

Figure \ref{fig:inv_a_dep_all_channels} shows a selection of invariant functions of leading twist based solely on data from $C_1$ contractions. The $A_{q\delta q'}$ signal is very similar to the one in $A_{\delta q q^\prime}$ while the $A_{\Delta q \Delta q^\prime}$ signal is primarily dominated by noise and thus not presented. We observe no clear $a$-dependence in the vector-vector and tensor-tensor channels, as data from the coarsest and finest ensembles studied coincide within $1\sigma$. In the tensor-vector channel, where we used \eqref{eq:dpd_sym} to enhance the signal in the $uu$ case, we observe a slight lattice spacing dependence. 

\subsubsection{$C_2$ contributions}
\begin{figure}[ht]
\centering

%================================================
% Row 1: VV
%================================================
\begin{subfigure}{0.49\textwidth}
\includegraphics[width=\textwidth]{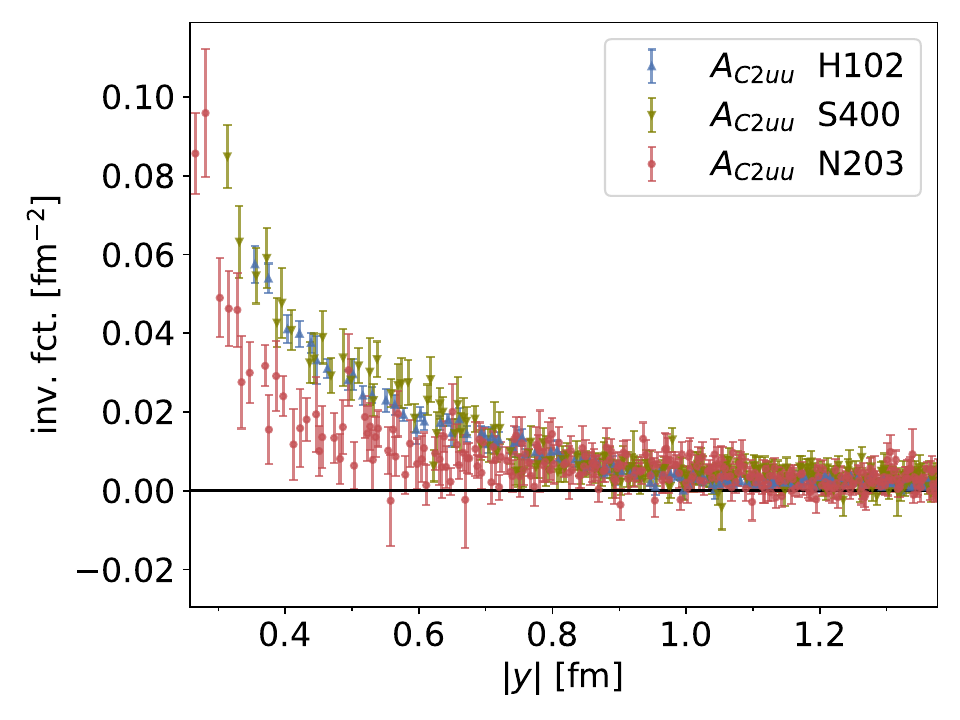}
\caption{$\mathrm{VV}$, $C_{2u}$}
\label{fig:inv_a_dep_VV_C2u}
\end{subfigure}
\hfill
\begin{subfigure}{0.49\textwidth}
\includegraphics[width=\textwidth]{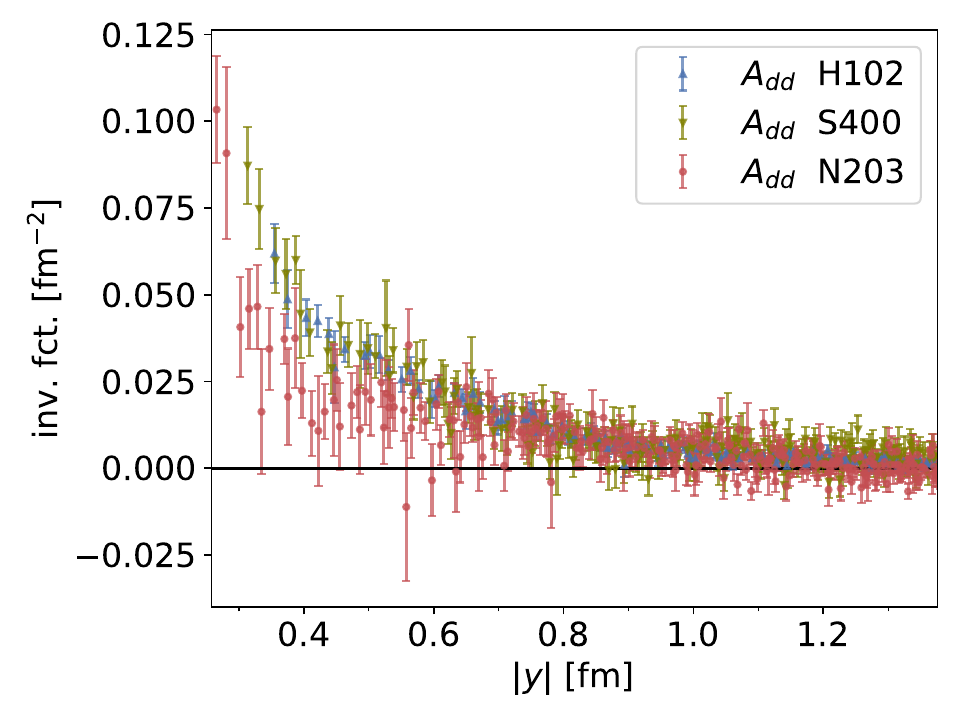}
\caption{$\mathrm{VV}$, $dd$}
\label{fig:inv_a_dep_VV_dd}
\end{subfigure}

\caption{
Leading twist invariant functions in the vector-vector channel for the flavor combinations $uu$, considering only the  $C_{2u}$ contraction (left), and $dd$ (right). Results are shown for the H102 ($a=0.085\,\mathrm{fm}$), S400 ($a=0.075\,\mathrm{fm}$), and N203 ($a=0.064\,\mathrm{fm}$) ensembles at $\vec{p}=(0,0,0)$ and $\cos\theta>0.9$.}
\label{fig:inv_a_dep_VV_TV_C2u_dd}
\end{figure}

Similarly to the previous subsection, Figure \ref{fig:inv_a_dep_VV_TV_C2u_dd} contains the flavor combinations $uu$ and $dd$, but obtained from data of $C_2$ contractions only. In the $A_{q q^\prime}$ data (top row), one can observe a dependence on the lattice spacing for small distances. For the finest lattice spacing $a=0.064\,\mathrm{fm}$, we find values smaller by up to around $2\sigma$ compared to the results obtained from our coarsest lattice with $a=0.085\,\mathrm{fm}$, in particular for $A_{C_{2u}}$ (Figure \ref{fig:inv_a_dep_VV_C2u}) and $A_{C_{2},dd}$ (Figure \ref{fig:inv_a_dep_VV_dd}). In the case of other channels, like $A_{C_2,\delta qq}$, the dependence on lattice spacing is less significant and uncertainties are larger, preventing a closer inspection.

\subsubsection{Flavor specific combinations}

\begin{figure}[ht]
\centering

%================================================
% Row 1
%================================================
\begin{subfigure}{0.48\textwidth}
\includegraphics[width=\textwidth]{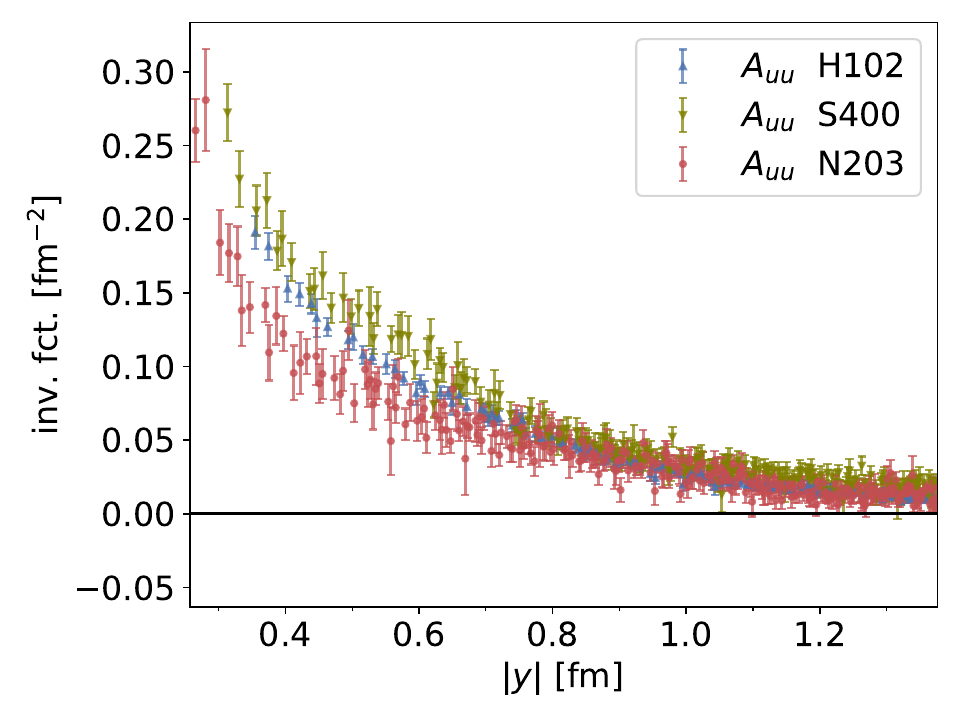}
\caption{$\mathrm{VV}$, $uu$}
\label{fig:inv_a_dep_VV_uu}
\end{subfigure}
\hfill
\begin{subfigure}{0.48\textwidth}
\includegraphics[width=\textwidth]{invfctcomp/Invplt_H102_S400_N203_VV_ud_p000_cos_0.9.pdf}
\caption{$\mathrm{VV}$, $ud$}
\label{fig:inv_a_dep_VV_ud_2x2}
\end{subfigure}

\vspace{0.4cm}

%================================================
% Row 2
%================================================
\begin{subfigure}{0.48\textwidth}
\includegraphics[width=\textwidth]{invfctcomp/Invplt_H102_S400_N203_VV_dd_p000_cos_0.9.pdf}
\caption{$\mathrm{VV}$, $dd$}
\label{fig:inv_a_dep_VV_dd_2x2}
\end{subfigure}
\hfill
\begin{subfigure}{0.48\textwidth}
\includegraphics[width=\textwidth]{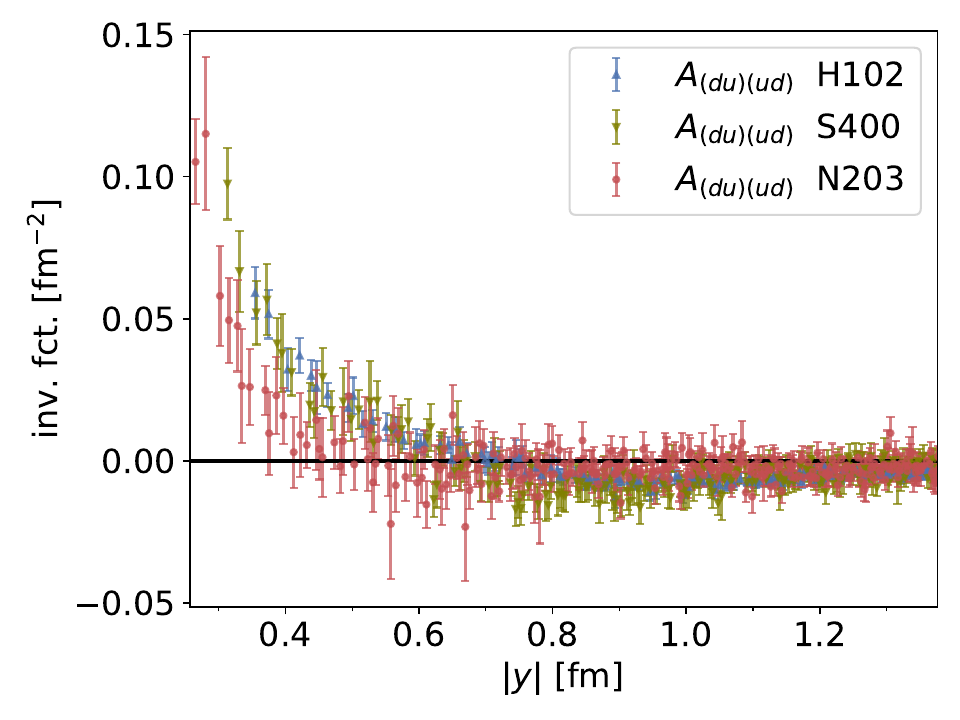}
\caption{$\mathrm{VV}$, $duud$}
\label{fig:inv_a_dep_VV_duud_2x2}
\end{subfigure}

\caption{
Leading twist invariant functions in the $\mathrm{VV}$ channel
for the flavor combinations $uu$, $ud$, $dd$, and $duud$
(arranged left to right, top to bottom).
Results are shown for the H102 ($a=0.085\,\mathrm{fm}$), S400 ($a=0.075\,\mathrm{fm}$), and N203 ($a=0.064\,\mathrm{fm}$) ensembles at 
$\vec{p}=(0,0,0)$ and $\cos\theta>0.9$.
}
\label{fig:inv_a_dep_VV_flavors}
\end{figure}

The resulting leading twist invariant functions $A_{qq^\prime}$ obtained from physical data combinations are presented in Figure \ref{fig:inv_a_dep_VV_flavors}. Both the residual finite volume effects from the $C_1$ contraction data on S400 as well as the lattice spacing dependence of the $C_2$ contraction are well visible in Figure \ref{fig:inv_a_dep_VV_uu} for the $uu$ flavor combination. The $ud$ flavor combination in Figure \ref{fig:inv_a_dep_VV_ud_2x2} is the same as in the $C_1$ contraction only case. Likewise, the $dd$ combination has no $C_1$ contribution and is thus the same as in Figure \ref{fig:inv_a_dep_VV_TV_C2u_dd}. The flavor interference combination $duud$ in Figure \ref{fig:inv_a_dep_VV_duud_2x2} exhibits no observable lattice spacing dependence within errors. It is unclear if the zero-crossing observed for H102 \cite{Reitinger:2024ulw} remains after taking the continuum limit, since the larger statistical error prevents us from drawing any conclusions in this direction.

\subsection{Mass dependence}
\label{sec:mdep}
In this section we investigate the meson mass dependence of the leading twist invariant functions. We adhere to the data selection criteria established in Section \ref{sec:artifacts}, namely that the angle $\theta$ enclosed by the separations with the closest lattice diagonal has to fulfill $\cos{\theta}>0.9$ and $4a\leq|\vec{y}|\leq La/2$. Just as for the lattice spacing dependence, we will first discuss the results from single contractions. Finally, we give a comparison of all physical flavor combinations. 
\subsubsection{$C_1$ contributions}

\begin{figure}[ht]
\centering

%================================================
% Row 1: VV (includes S100)
%================================================
\begin{subfigure}{0.49\textwidth}
\includegraphics[width=\textwidth]{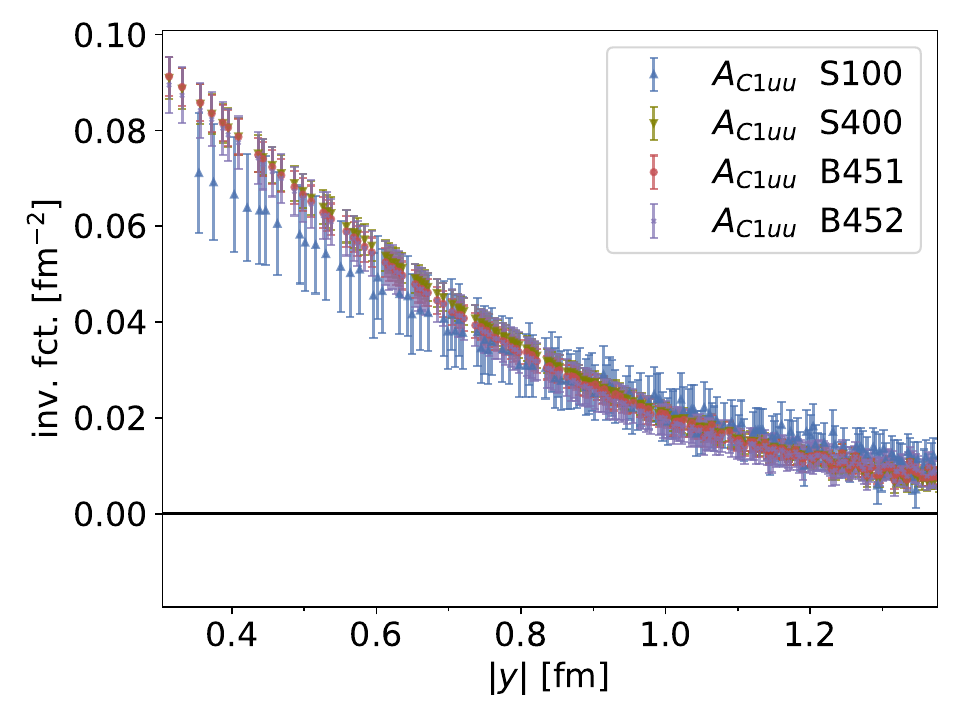}
\caption{$\mathrm{VV}$, $C_{1,uuuu}$}
\label{fig:inv_m_dep_VV_uuuu}
\end{subfigure}
\hfill
\begin{subfigure}{0.49\textwidth}
\includegraphics[width=\textwidth]{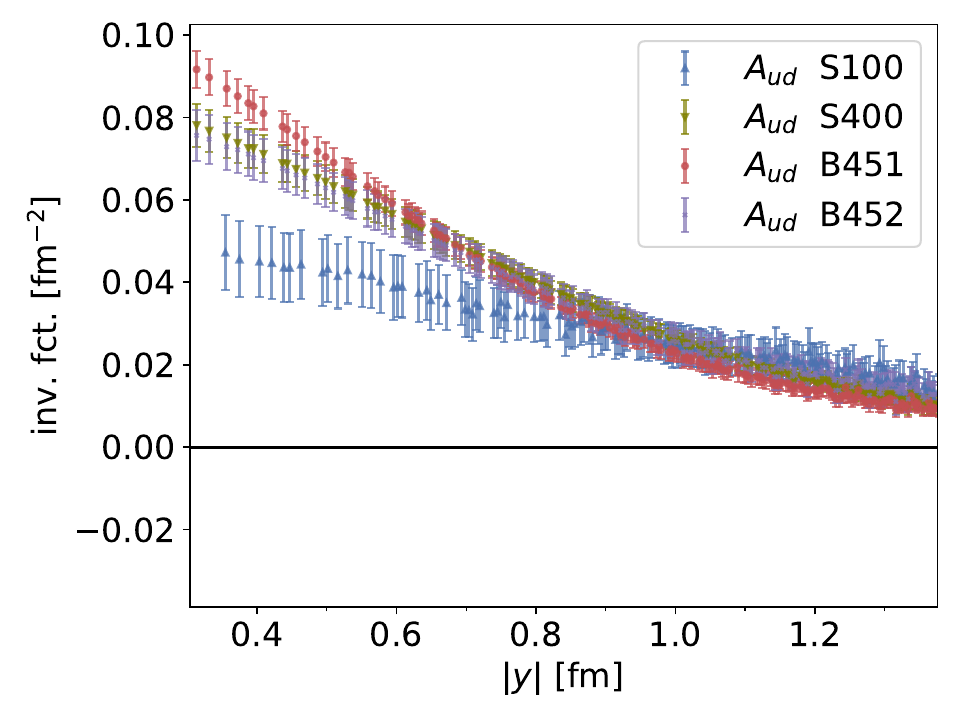}
\caption{$\mathrm{VV}$, $C_{1,ud}$}
\label{fig:inv_m_dep_VV_ud}
\end{subfigure}

\vspace{0.4cm}

%================================================
% Row 2: TV
%================================================
\begin{subfigure}{0.49\textwidth}
\includegraphics[width=\textwidth]{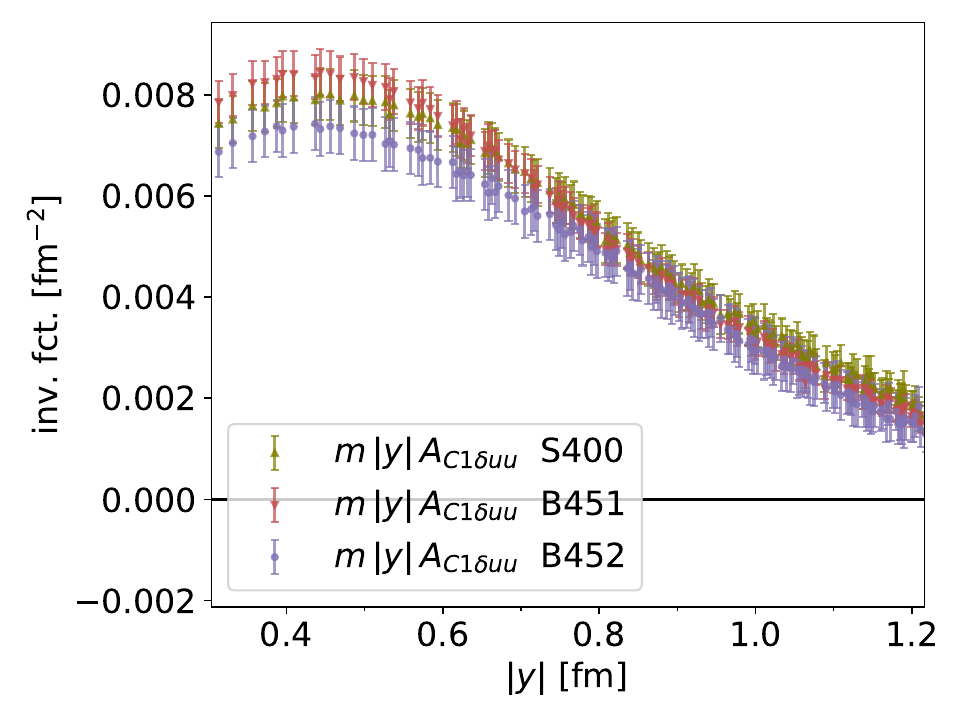}
\caption{$\mathrm{TV}$, $C_{1,uuuu}$}
\label{fig:inv_m_dep_TV_uuuu}
\end{subfigure}
\hfill
\begin{subfigure}{0.49\textwidth}
\includegraphics[width=\textwidth]{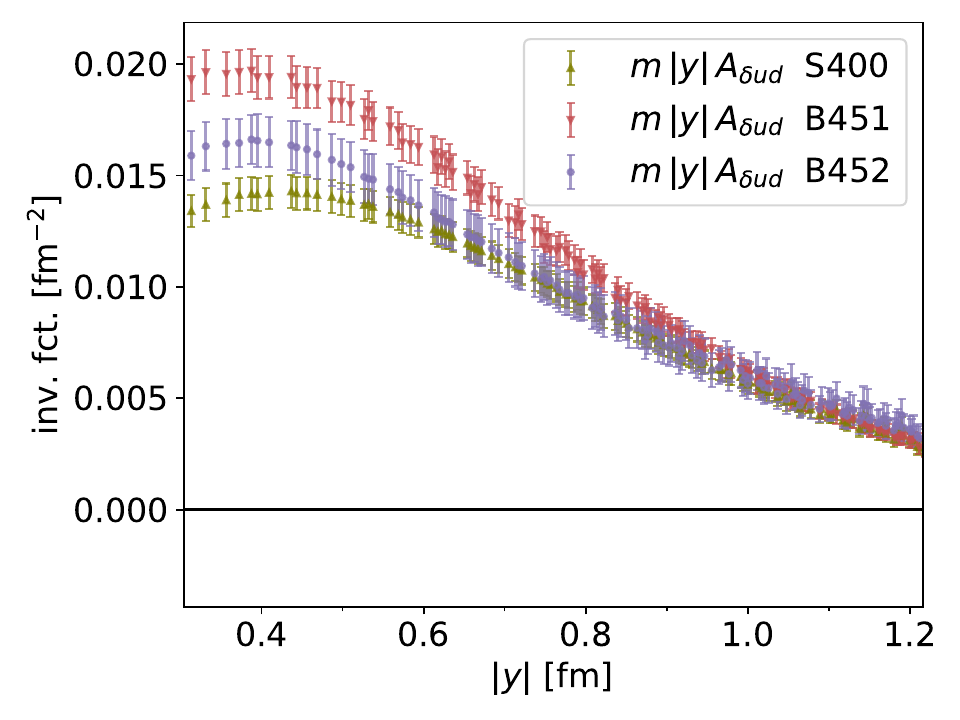}
\caption{$\mathrm{TV}$, $C_{1,ud}$}
\label{fig:inv_m_dep_TV_ud}
\end{subfigure}

\caption{
Leading twist invariant functions illustrating the mass dependence in the vector-vector (top row) and tensor-vector (bottom row) channel for the flavor combinations $C_{1,uuuu}$ (left column) and $C_{1,ud}$ (right column). Results are displayed at $\vec{p}=(0,0,0)$ and $\cos\theta>0.9$, including only data from the $C_1$ contraction. The ensembles S100 ($m_\pi=214\,\mathrm{MeV}$), S400 ($m_\pi=354\,\mathrm{MeV}$), B451 ($m_\pi=422\,\mathrm{MeV}$), and B452 ($m_\pi=352\,\mathrm{MeV}$) are used to probe the pion-mass dependence. For the vector-vector channel, data from all four ensembles are shown, while for the tensor-vector channel, the comparison includes S400, B451, and B452.}
\label{fig:inv_m_dep}
\end{figure}

Figure \ref{fig:inv_m_dep} contains the leading twist invariant functions $A_{qq^\prime}$ and $A_{\delta qq^\prime}$ extracted from $C_1$ contraction data for all flavors. It is structured the same way as for the lattice spacing dependence case. For $A_{qq^\prime}$ (first) row, data obtained from the S100 ensemble is presented additionally, where the pion mass is $m_\pi = 214\,\mathrm{MeV}$. Notice that the lattice spacing of S100 is coarser than that of the other ensembles used in this plot. This is not an issue, since there is no significant dependence on the lattice spacing in the case of $A_{qq}$ (see subsection \ref{subsec:a_dep_C1}). 
 
We find  a definite mass dependence in multiple cases, e.g.\ in $A_{ud}$ (Figure \ref{fig:inv_m_dep_VV_ud}) and $A_{duud}$ (not shown), consistently following the trend of reducing the absolute value of the signal as the light quark masses decrease. This decrease is most prominent when comparing data of ensembles of different pion mass in descending order, here B451, S400 and S100. While the signals from the S400 and B452 data, which differ only in strange quark mass, agree in most cases, distinct differences $\approx2\sigma$ can be observed in $A_{\delta ud}$ (Figure \ref{fig:inv_m_dep_TV_ud}) and $A_{\delta(du)ud}$ (not shown), which may indicate a possible dependence on the strange sea-quark mass, although the significance is not sufficient to establish such an effect. For the flavor combination $uu$, we find no mass dependence in any channel.

\subsubsection{$C_2$ contributions}
\begin{figure}[ht]
\centering

%================================================
% Row 1: VV
%================================================
\begin{subfigure}{0.48\textwidth}
\includegraphics[width=\textwidth]{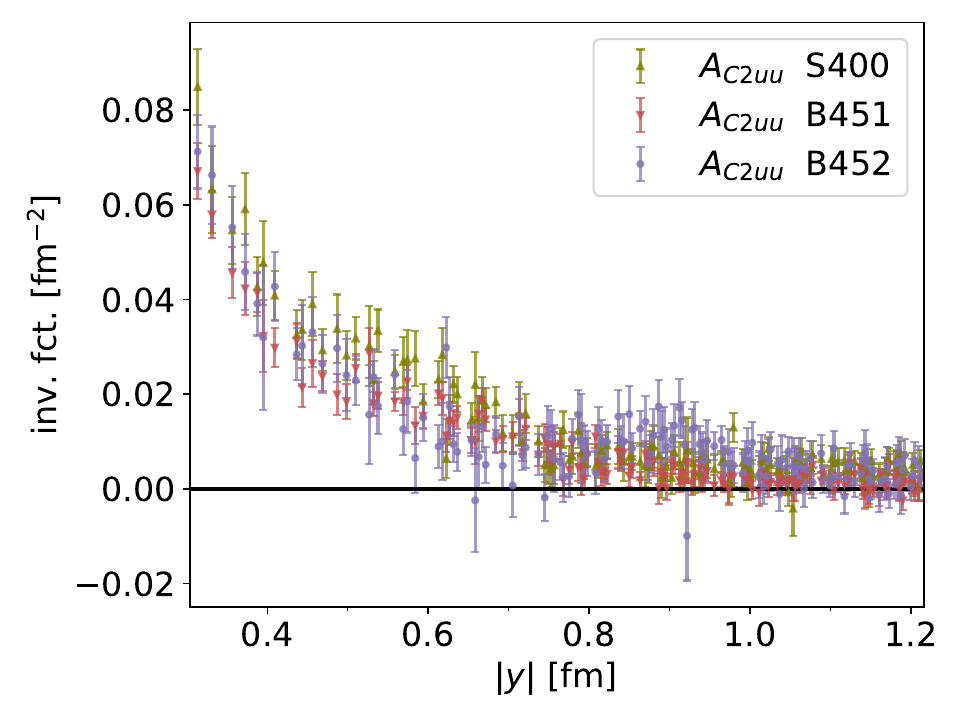}
\caption{$\mathrm{VV}$, $C_{2u}$}
\label{fig:inv_m_dep_VV_C2u}
\end{subfigure}
\hfill
\begin{subfigure}{0.48\textwidth}
\includegraphics[width=\textwidth]{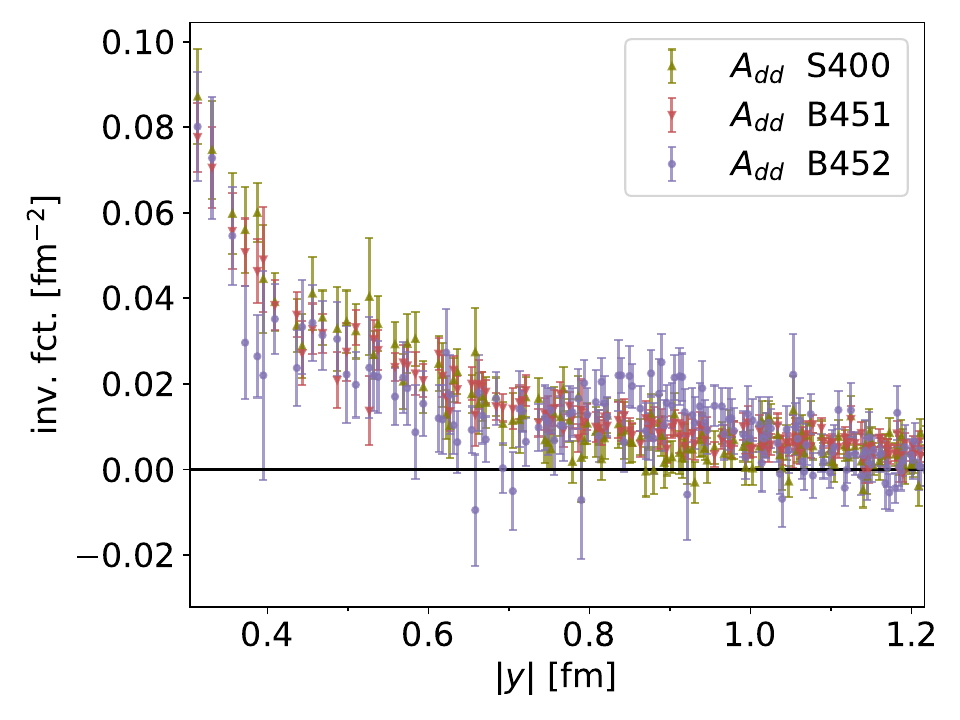}
\caption{$\mathrm{VV}$, $dd$}
\label{fig:inv_m_dep_VV_dd}
\end{subfigure}
\caption{
Leading twist invariant functions illustrating the mass dependence in the vector-vector channel for the flavor combinations $C_{2u}$ (left) and $dd$ (right). Results are displayed for the S400 ($m_\pi=354\,\mathrm{MeV}$), B451 ($m_\pi=422\,\mathrm{MeV}$), and B452 ($m_\pi=352\,\mathrm{MeV}$) ensembles at $\vec{p}=(0,0,0)$ and $\cos\theta>0.9$.}
\label{fig:inv_m_dep_VV_TV_C2u_dd}
\end{figure}

The leading twist invariant functions extracted form $C_2$ contraction exhibit no mass dependence in any channel. As an example, $A_{qq^\prime}$ is depicted in Figure \ref{fig:inv_m_dep_VV_TV_C2u_dd}. As in the lattice spacing case, $A_{\Delta q \Delta q^\prime}$ and $A_{\delta q \delta q^\prime}$ data is too beset with noise as to lead to any sensible conclusions and is thus excluded.

\subsubsection{Flavor specific combinations}

\begin{figure}[ht]
\centering

%================================================
% Row 1
%================================================
\begin{subfigure}{0.48\textwidth}
\includegraphics[width=\textwidth]{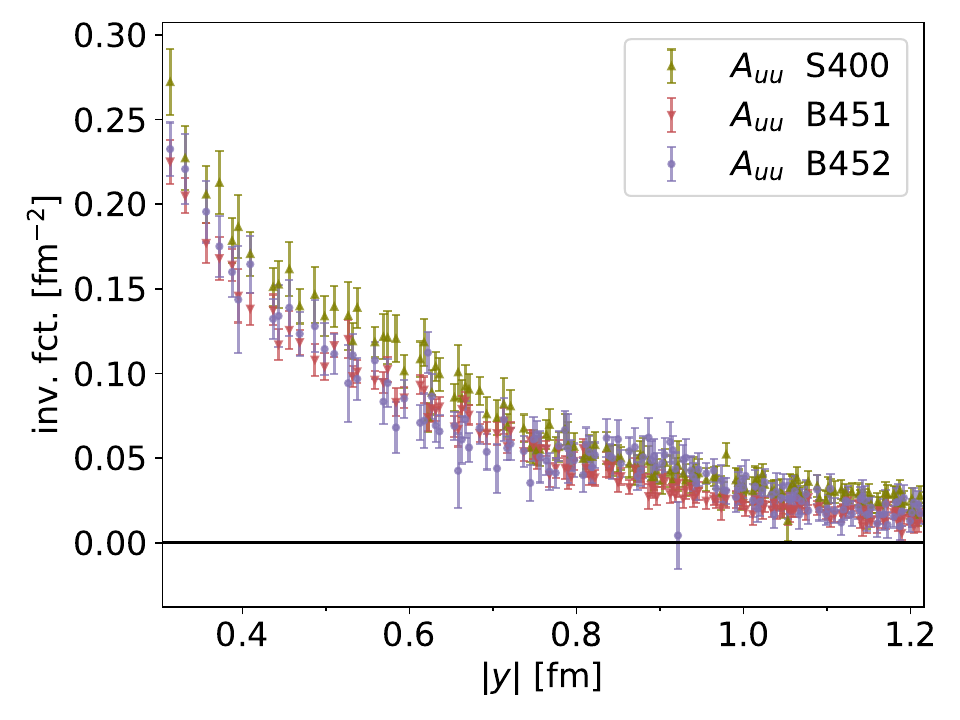}
\caption{$\mathrm{VV}$, $uu$}
\label{fig:inv_m_dep_VV_uu}
\end{subfigure}
\hfill
\begin{subfigure}{0.48\textwidth}
\includegraphics[width=\textwidth]{invfctcomp/Invplt_S100_S400_B451_B452_VV_ud_p000_cos_0.9.pdf}
\caption{$\mathrm{VV}$, $ud$}
\label{fig:inv_m_dep_VV_ud_2x2}
\end{subfigure}

\vspace{0.4cm}

%================================================
% Row 2
%================================================
\begin{subfigure}{0.48\textwidth}
\includegraphics[width=\textwidth]{invfctcomp/Invplt_S400_B451_B452_VV_dd_p000_cos_0.9.pdf}
\caption{$\mathrm{VV}$, $dd$}
\label{fig:inv_m_dep_VV_dd_2x2}
\end{subfigure}
\hfill
\begin{subfigure}{0.48\textwidth}
\includegraphics[width=\textwidth]{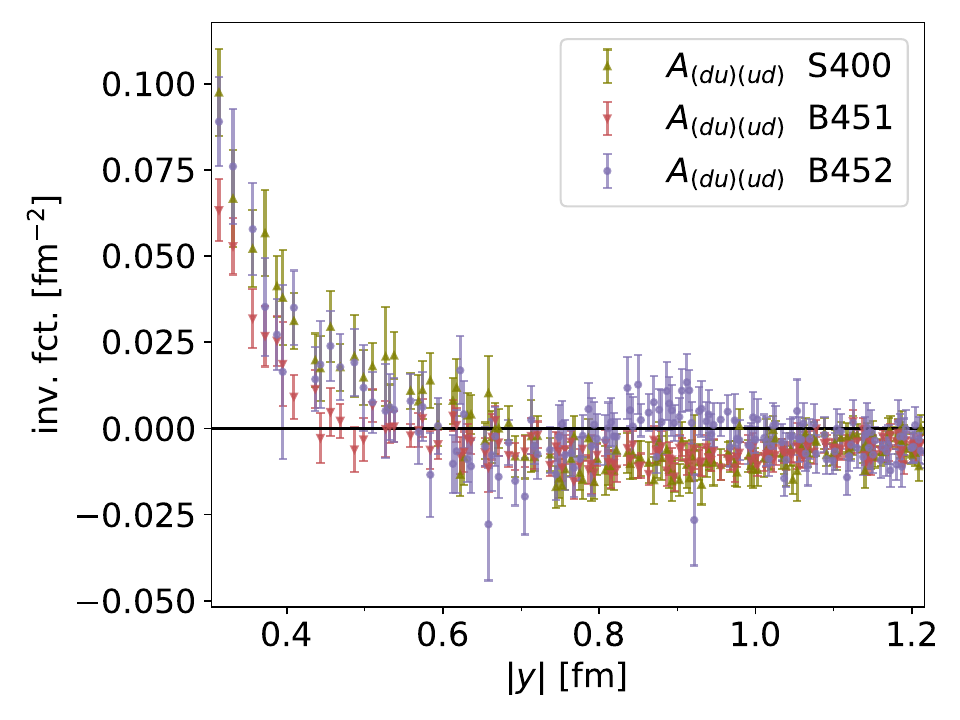}
\caption{$\mathrm{VV}$, $duud$}
\label{fig:inv_m_dep_VV_duud_2x2}
\end{subfigure}

\caption{
Leading twist invariant functions illustrating the mass dependence
in the $\mathrm{VV}$ channel for the flavor combinations
$uu$, $ud$, $dd$, and $duud$
(arranged left to right, top to bottom).
Results are shown at $\vec{p}=(0,0,0)$ and $\cos\theta>0.9$.
For $uu$, $dd$, and $duud$ the ensembles S400 ($m_\pi=354\,\mathrm{MeV}$), B451 ($m_\pi=422\,\mathrm{MeV}$), and B452 ($m_\pi=352\,\mathrm{MeV}$) are included,
while for the $ud$ case additional data from the S100 ($m_\pi=214\,\mathrm{MeV}$) ensemble are shown.
}
\label{fig:inv_m_dep_VV_flavors}
\end{figure}

The resulting leading twist invariant functions $A_{qq^\prime}$ obtained from physical data combinations are presented in Figure \ref{fig:inv_m_dep_VV_flavors}. The $uu$ and $dd$ combinations exhibit no observable mass dependence. The $ud$ flavor combination data is the same as in Figure \ref{fig:inv_m_dep_VV_ud}, as it contains no $C_2$ contribution, exhibiting a clear mass dependence. All other combinations include contributions from $C_2$ contractions, which exhibit a definite $a$-dependence, thus the S100 data is not included in their comparison. We find no evidence for mass dependence in  any channel for the combinations $uu$, $dd$ and $duud$.

\section{Summary}
\label{sec:summary}

We have investigated the dependence of double parton distribution observables
on the pion and kaon masses as well as the lattice spacing, using six CLS
gauge ensembles with pion masses ranging from $214\,\mathrm{MeV}$ to
$422\,\mathrm{MeV}$ and lattice spacings between $0.064\,\mathrm{fm}$ and
$0.085\,\mathrm{fm}$.\\

Prior to the main analysis, we examined potential sources of systematic uncertainty. Excited state contamination was assessed by comparing results at multiple source-sink separations on the S100 and S400 ensembles, finding the plateaus to be consistent within statistical errors. We investigated finite volume effects and discretization artifacts through an anisotropy analysis of the matrix elements and derived invariant functions. We find that after applying the cut of $\cos{\theta}>0.9$, finite volume effects at large distances are sufficiently under control.

We find no significant lattice spacing dependence in the $C_1$ contraction data. The $C_2$ contraction exhibits a moderate lattice spacing dependence at small separations in the vector-vector channel, while other channels appear to be less affected or too noisy to be assessed. Overall, the lattice spacing dependence of the DPD observables studied here is mild, suggesting that simulations at currently accessible lattice spacings already provide physically meaningful results.\\

In our investigation of the light and strange quark mass dependence, the $C_2$ contraction data exhibited no significant dependence on the pion or kaon mass. In the $C_1$ contraction data, a clear mass dependence is observed in several invariant functions, in particular $A_{ud}$, $A_{duud}$, and $A_{\delta ud}$, where the signal decreases as the light quark mass is reduced. A comparison of the S400 ($m_\pi=354\,\mathrm{MeV}, m_K=445\,\mathrm{MeV}$) and B452 ($m_\pi=352\,\mathrm{MeV}, m_K=548\,\mathrm{MeV}$) ensembles, which share the same pion mass but differ in the strange quark mass, reveals differences of up to $2\sigma$ in selected channels, hinting at potential sea quark effects. Given that a mass dependence is observed even well above the physical point, simulations closer to the physical pion mass will be necessary for a quantitative extrapolation. The non-vanishing signal at the lowest pion mass studied, $m_\pi = 214\,\mathrm{MeV}$, confirms that DPD-related observables remain accessible on the lattice at near-physical quark masses.

\acknowledgments
We thank M. Diehl for fruitful discussions.
DR and AS are supported by the Deutsche Forschungsgemeinschaft  (DFG) grant SCHA 458/23, project number 493441321.
CZ is partially supported by DOE Quark-Gluon Tomography (QGT) Topical Collaboration under award No. DE-SC0023646. CZ is supported in part by the Alexander von Humboldt Foundation and the U.S. Department of Energy, Office of Science, Office of Nuclear Physics, under Grant No. DE-SC0013065 and under Contract No. DE-AC02-05CH11231 which is used to operate Lawrence Berkeley National Laboratory.
This research used resources of the National Energy Research Scientific Computing Center (NERSC), a U.S. Department of Energy Office of Science User Facility located at Lawrence Berkeley National Laboratory, operated under Contract No. DE-AC02-05CH11231.
The authors gratefully acknowledge the scientific support and HPC resources provided by the Erlangen National High Performance Computing Center (NHR@FAU) of the Friedrich-Alexander-Universität Erlangen-Nürnberg (FAU) under the NHR project b164da. NHR funding is provided by federal and Bavarian state authorities. NHR@FAU hardware is partially funded by the German Research Foundation (DFG) – 440719683.
We acknowledge the CLS effort for generating the $n_f = 2 + 1$ ensembles in
\cite{Bruno:2014jqa}, which were used for this work. 
%\newpage
\phantomsection
\addcontentsline{toc}{section}{References}
\bibliographystyle{JHEP}
\bibliography{DPD_mass_paper}

\end{document}